\documentclass[letterpaper, 10pt]{IEEEtran}

\IEEEoverridecommandlockouts

\usepackage{calrsfs,times}
\usepackage{graphicx,psfrag,enumerate}
\usepackage{amssymb}
\usepackage{amsmath}
\usepackage{calrsfs}
\usepackage{color, epsfig, epstopdf, wrapfig}
\usepackage{paralist}
\usepackage{float,cuted}
\usepackage{subcaption}

\newtheorem{definition}{Definition}
\newtheorem{theorem}{Theorem}
\newtheorem{remark}{Remark}

\newtheorem{lemma}{Lemma}

\usepackage{amsmath}
\usepackage{accents}
\newlength{\dhatheight}

\usepackage{color}
\usepackage[normalem]{ulem}
\usepackage{soul} 
\soulregister\cite7
\soulregister\ref7
\soulregister\pageref7

\begin{document}

\title{\LARGE \bf 
  Distributed $H_\infty$ Estimation Resilient to 
  Biasing Attacks\thanks{This work was supported by the Australian
  Research Council and the University of New South Wales.}}

\author{V.~Ugrinovskii\thanks{V. Ugrinovskii is with the School of Engineering and Information Technology, University of New South Wales Canberra, Canberra, ACT 2600, Australia. {\tt\small v.ougrinovski@adfa.edu.au}}}

\maketitle
         
\begin{abstract}
We consider the distributed $H_\infty$ estimation problem with an additional
requirement of resilience to biasing attacks. An attack scenario
is considered where an adversary misappropriates some of the observer nodes
and injects biasing signals into observer dynamics. The paper proposes
a procedure for the derivation of a distributed observer which 
endows each node with an
attack detector which also functions as an attack compensating feedback
controller 
for the main observer. Connecting these controlled 
observers into a network results in a distributed observer whose nodes
produce unbiased robust estimates of the plant. We show that the gains for
each controlled observer in the network can be computed in a decentralized
fashion, thus reducing vulnerability of the network.  
\end{abstract}

\section{Introduction}

The topic of understanding possible attacker's strategies and developing
suitable countermeasures receives considerable attention in the 
literature dedicated to designing control and filtering systems resilient
to malicious interference. The approaches of the
game and optimization theories are conventional in this
area~\cite{ACS-2009,GLB-2010,LaU3,ZCSC-2015,DT-2015,ZB-2015}, although 
they often tend to overestimate the attacker, assuming that the adversary  
has sufficient resources to collect detailed measurements and implement a
sophisticated attack policy. For instance, the recent Byzantine models of the
adversarial behavior allow the adversarial nodes to possess complete
knowledge about the graph topology and the plant dynamics, i.e., an
adversarial node knows the measurements received by the healthy
nodes~\cite{MS-2016}. This has undesirable consequences of having to endow the
network with a substantial redundancy, in order to guarantee that it is
able to sustain multiple simultaneous attacks~\cite{CHT-2018,FTD-2014,MS-2016}.

There have been suggestions that robust control techniques could provide an
alternative framework for optimizing the system performance in the
presence of an attacker~\cite{CRW-2014}. Indeed, robust control models are
quite flexible and can handle large numbers of
uncertainty inputs.
  In this regard they may offer a certain flexibility
over the Byzantine model in that they are generally less restrictive about
the number of nodes affected by the adversary. However, robustness
specifications differ significantly from requirements for resilience to
strategic attacks~\cite{ZB-2015}, which could make the robust control
approach too conservative as the means for accommodating strategic
intentions of a malicious attacker~\cite{FTD-2014}. 

In this paper we show that this drawback of the robustness methodology can
be overcome if the features of the attack which make it distinct from
benign disturbances are properly accounted. To this end, we consider the 
so-called biasing attacks~\cite{TSSJ-2015} which admit a simple mathematical
description, and yet capture a nontrivial type of misappropriation attacks
on distributed networks. To be concrete, the paper focuses on the
problem of resilient distributed estimation using sensor networks in the
presence of a strategic adversary who misappropriates a number of sensing
nodes and injects a biasing signal into the
observers~\cite{DUSL1,TSSJ-2015}. 
To interfere with operation of the network the intruder first
infiltrates the network, by breaching the cyber layer of security. Once the 
adversary has penetrated the cyber security barrier and has gained access
to the estimation algorithm at the misappropriated node, it modifies it by
injecting biasing signals. In this regard, our attack scenario resembles
that of the Stuxnet computer worm~\cite{Smith-2015}. The injected biasing
signals cause the misappropriated observers to produce erroneous estimates
of the plant.  Also, the bias spreads through the network interconnections
causing a `domino' effect of cascading errors, forcing the entire observer
network to fail. To achieve this, the attacker does not need any
knowledge of the system or the plant.

We show that information collected by sensors within a typical
network of state observers coupled in a diffusive manner~\cite{MFN-2013} 
is sufficient for discovering and disarming such biasing
misappropriation attacks. To demonstrate this, we  
propose a distributed observer architecture which augments the 
observer network with an additional layer consisting 
of a network of interconnected attack detecting controllers. These controllers
utilize the same noisy innovation information which is used for state
estimation. based on this information, they estimate an extended vector of
observer errors and use it for feedback. We show that this feedback control
policy has an asymptotically negligible effect on the dynamics of healthy
observers. However at the compromised nodes, the controller outputs track
the biasing attack inputs; this enables to negate the attack. The
controllers are constructed within the $H_\infty$ control framework. The
choice of the framework is consistent with the original state estimation
objective and facilitates the performance analysis of the resilient
estimation algorithm.

Controlling the plant observers via
feedback to make them resilient to the biasing behavior of misappropriated
nodes is the main distinction of this paper from our previous
work~\cite{DUSL1}. As another distinct feature, the paper considers the
most general biasing attack scenario in which the attacker seeks to bias
both the state observer and the attack detector. Indeed, the attacker who
has gained access to the data processing algorithm will likely to temper with
both the state observer and security devices. Hence, the attack detection
algorithm  must also be made resilient to attacks, in the same manner as
the plant observer. Although the paper considers this problem in the context
specific to misappropriation attacks on distributed observers, we believe
the problem is important in general since security devices cannot
not be assumed to be immune to malicious interference. 

Starting from the seminal work~\cite{Olfati-Saber-2007}, distributed
computations are widely used in the literature on distributed
estimation. In addition to solving the main task of obtaining state
estimates of the plant, in many algorithms, sensor nodes must communicate 
to compute parameters of their observers, such as error covariance
matrices, innovation gains, etc~\cite{Olfati-Saber-2007}. Such distributed
computations present obvious security risk, and hardening of distributed
algorithms is a subject of current research~\cite{SG-2016}. In this paper,
we approach this issue differently, aiming to achieve a computational
autonomy of the observers. Namely, in the proposed method, each observer
node computes its observer in a decentralized fashion, without
communicating with other nodes. 
The method is based on the distributed observer design approach proposed
in~\cite{ZU1a}. It involves an initial setup step during which certain
auxiliary parameters are computed which are then assigned to the nodes.
This initial step is carried out centrally when the network is
offline, hence it does not jeopardize the system security. Also, the
knowledge of the plant observed is not required at this stage, since these
parameters are computed only using characteristics of the communication
network and the desired performance characteristics. Each sensor node then
utilizes these auxiliary parameters  to compute its node  
observer autonomously. The plant model must be used at this stage, of course.  
This methodology contrasts the methodology used in our previous
work~\cite{DUSL1} which 
relied on solving certain linear matrix inequalities coupled among the nodes to compute
attack detectors.

In summary, the main contribution and features of the paper are as follows:
\begin{enumerate}[(a)]
\item 
We introduce a new class of networked attack detectors that use feedback to
suppress biasing attacks on distributed observer networks. 

\item
Despite the proposed methodology of resilient estimation 
is based on the approach of $H_\infty$ control, it 
distinguishes between benign disturbances and attack signals. 

\item
Formally, our approach does not limit the number of
nodes subjected to simultaneous biasing attacks. Essentially, every node is
treated conservatively as a potential target. However, trusted nodes can be easily accommodated in
our formulation. This will only simplify the system model used in the derivation of the attack detector and controller.  

\item
The proposed methodology identifies the nodes subjected to biasing attacks,
regardless whether the adversary targets the state observer or its attack
detecting feedback controller, or both. 

\item
The node observers compute their parameters in decentralized manner in real
time. Only the information about their states and states of their
controllers needs to be shared through communication channels. 

\item
In the special case of an LTI plant and observers, we are able to provide a
deeper insight into feasibility and performance optimization of the
proposed resilient observers, including selecting the best network
topology from a set of candidate topologies. 
\end{enumerate}

The preliminary version of the paper has been presented at
the 2018 American Control Conference~\cite{U10b}. Compared with the conference
version, the current version contains a substantial amount of new material
concerned with security of the proposed attack detecting controllers and
performance optimization over network graphs. Also, the paper includes 
an illustrating example which was left out from the conference version due
to space constraints. 

The paper begins with presenting a background on distributed
filtering in Section~\ref{sec:distributed_estimation}. Also, in that
section we describe the class of biasing attacks which will be used in the
derivation of our main results. The formal problem statement is given in 
Section~\ref{Problem.formulation}. The procedure to construct a resilient
distributed observer is described in Section~\ref{resilient.estimation},
and its feasibility and optimization aspects are discussed in
Section~\ref{opt}. It 
is illustrated in Section~\ref{sec:simulations}
with an example. The conclusions are given in Section~\ref{sec:conclusion}.

\emph{Notation}: $\mathbf{R}^n$ denotes the real Euclidean $n$-dimensional vector space, with the norm  $\|x\|=(x'x)^{1/2}$; here the symbol $'$ denotes the transpose of a matrix or a vector.
The symbol $I$ denotes the identity matrix. For real symmetric
$n\times n$ matrices $X$ and $Y$, $Y>X$ (respectively, $Y\geq X$) means the
matrix $Y-X$ is positive definite (respectively, positive
semidefinite). $\otimes$ is the Kronecker product of matrices.
$\|x\|_X$ denotes the weighted norm of $x$: $\|x\|_X=(x'Xx)^{1/2}$.  
$\mathrm{diag}\left[C_1,\ldots, C_N\right]$ denotes the block-diagonal matrix
with the matrices $C_1$, \ldots, $C_N$ as its diagonal blocks. 
The notation $L_2[0, \infty)$ refers to the Lebesgue space of
$\mathbf{R}^n$-valued vector-functions $z(.)$, defined on the time interval
$[0, \infty)$, with the norm $\|z\|_2\triangleq\left(\int_0^\infty
  \|z(t)\|^2 dt \right)^{1/2}$ and the inner product $\int_0^\infty z_1'(t)
z_2(t) dt$. For a causal signal $f(t)$, $f(s)$ denotes the Laplace
transform of $f(t)$.  
   
\section{Biasing misappropriation attacks on distributed observers and
  resilient distributed estimation}
\label{sec:distributed_estimation}

\subsection{Biasing misappropriation attacks on observer networks}

Consider a linear time-varying plant 
\begin{eqnarray}
\label{state}\label{eq:plant}
 \dot{x} = A(t)x +B(t)w, \quad x(0)=x_0,
\end{eqnarray}
subject to an unknown disturbance $w\in L_2[0,\infty)$. The plant is
observed by a 
network of $N$ sensors. The sensor at node $i$ collects measurements of the
plant, corrupted by disturbances $v_i$: 
\begin{equation}\label{measurement}
y_i = C_i(t)x + D_i(t)v_i, \quad y_i\in\mathbf{R}^{p_i}.
\end{equation}
It also exchanges information with other nodes. The 
communication network forms a directed graph $\mathbf{G}$. Without loss of
generality, the graph is
assumed to be connected but not necessarily strongly
connected\footnote{
  This assumption is justified due to Proposition~1 in~\cite{U6}.
  The sufficiency part of that proposition can be easily extended to
  distributed estimation problems with a general cost considered in this
  paper. It implies that if the network is disconnected, to obtain a
  solution to the distributed $H_\infty$ filtering problem over such
  network, it suffices to obtain a solution to the distributed $H_\infty$
  filtering problem over each connected network component.\label{Comment2-conn}}.
For each node $i$, let 
$\mathbf{V}_i$ denotes the set of its neighbors supplying information
to that node. The information received by node $i$ from
its neighbor $j\in \mathbf{V}_i$ is a $p_{ij}$-dimensional signal  
\begin{equation}\label{communication}
c_{ij} = W_{ij}\hat{x}_j + H_{ij}v_{ij},
\end{equation}
which contains information about the neighbor $j$'s estimate
$\hat x_j$ of the plant state $x$. The $p_{ij}\times n$ matrix $W_{ij}$
determines what information about $\hat x_j(t)$
node $j$ shares with node $i$. The signal $v_{ij}\in L_2[0,\infty)$ in 
(\ref{communication}) represents a channel disturbance. We think of the
matrices $W_{ij}$, $H_{ij}$ as characteristics of the network which is
considered to be fixed (the motivation for this will be explained later)
and independent of the plant. For that reason, these matrices  are assumed
to be constant; cf.~\cite{MS-2018}.      

We now make standing assumptions about the coefficients of the system
(\ref{eq:plant}), (\ref{measurement}), (\ref{communication}). Throughout
the paper, it will be assumed that the matrix-valued functions $A(t)$,
$B(t)$, $C_i(t)$, $D_i(t)$, $i=1, \ldots, N$, are bounded on the interval
$[0,\infty)$. Also, it will be assumed that $(D_i(t)D_i(t)')^{-1}$ exists
and is bounded on 
$[0,\infty)$ for all $i$. 

In the distributed
estimation scenario, the measurements and the communicated information are 
processed at the sensor nodes rather than centrally. Following~\cite{Olfati-Saber-2007}, 
the following observers are often considered for this~\cite{MK-2013,U6,SS-2009,HTWS-2018}:
\begin{eqnarray}
\dot{\hat{x}}_i &=& A(t)\hat x_i + L_i(t)(y_i-C_i(t)\hat x_i) \nonumber \\
&&+\sum_{j\in\mathbf{V}_i}K_{ij}(t)(c_{ij}-W_{ij}\hat x_i), \quad
\hat{x}_i(0)=\xi_i.
\label{filter_i}\label{UP7.C.d.unbiased}
  \end{eqnarray}
Since the plant is time-varying, 
the coefficients
$L_i$, $K_{ij}$ in (\ref{UP7.C.d.unbiased}) are assumed to be time-varying
in general.   
The observers (\ref{UP7.C.d.unbiased}) are coupled. The 
coupling is especially useful when some of the pairs
$(A(t),C_i(t))$ are not detectable\footnote{Here we follow
    the definition of detectability in~\cite{NK91} and say that
    the pair  $(A(t), C_i(t))$ is
  detectable if there exists a gain $\tilde L(t)$ such that the system
  $\dot e=(A(t)-\tilde L(t)C_i(t))e$ is exponentially stable.}. In this
case, the corresponding nodes
  face the situation that their filters may not be able to track the plant
if they rely on local measurements alone. 
The interconnections provide those nodes
with an information about $x(t)$ which cannot be obtained  
from the local measurement $y_i$ but can be extracted from the
neighbors' messages
$c_{ij}$~\cite{U6,U7b-journal,MS-2018}. Distributed observers
of the form (\ref{filter_i}) of course require the system to be detectable
  as whole; i.e., the pair $(A(t),[C_1',\ldots, C_N']')$ must be detectable.


A common problem of distributed estimation is to ensure that
the node estimates $\hat x_i(t)$ track $x(t)$ (or a part thereof) with
acceptable accuracy and robustness against disturbances in the plant model, measurements and interconnection 
channels; e.g., see~\cite{Olfati-Saber-2007,PM-2017,U6}.
However, the dependency on information sharing
makes a network of observers (\ref{UP7.C.d.unbiased}) vulnerable to
attacks seeking to disrupt this task. Common scenarios of such attacks 
involve an adversary injecting false signals into sensor
measurements or communicated data~\cite{PDB-2013} or behaving as a
Byzantine fault~\cite{MS-2016}. The latter behavior assumes that the adversary 
can misappropriate some network nodes and force them to deviate arbitrarily
from the prescribed estimation algorithm and transmit different false state
estimates to different neighbors. In this paper, we consider a similar
adversarial behavior, however we assume that the adversary 
strategically substitutes the node observers
(\ref{UP7.C.d.unbiased}) with their biased versions~\cite{DUSL1},  
\begin{eqnarray}  
    \dot{\hat x}_i&=&A(t)\hat x_i + L_i(t)(y_i(t)-C_i(t)\hat x_i) \nonumber \\
&& +\sum_{j\in
      \mathbf{V}_i}K_{ij}(t)(c_{ij}-W_{ij}\hat x_i)+F_if_i, \quad \hat
    x_i(0)=\xi_i,\quad 
  \label{UP7.C.d}
\end{eqnarray}
Here $f_i\in \mathbf{R}^{n_{f_i}}$ is an unknown signal representing the
attacker's input.

When node $i$ is forced to use the observer 
(\ref{UP7.C.d}) in lieu of (\ref{UP7.C.d.unbiased}), it generates biased
estimates of $x(t)$ closely resembling those that could be obtained using 
the true observer (\ref{UP7.C.d.unbiased}). These biased
estimates $\hat x_i$ are then broadcast across the network and will bias
other nodes. To disrupt the network, the adversary does not need to know
the plant dynamics, the measurements or the communication graph of the system;
cf.~\cite{MS-2016}. Different from the Byzantine attack
modeling, we will not have to impose a limit on the number of
misappropriated nodes, and can consider the worst-case situation where
every node of the observer network can be biased.

\begin{remark}\label{Comment2-2}
Biasing attacks of this type can occur as a result of a strategic network 
intrusion, when the adversary breaches the cyber security layer and gains
access to the observer algorithm at the 
misappropriated nodes. In this sense, our attack model is conceptually
different from models of benign disturbances.  Also unlike benign
disturbances, in our model the biasing inputs disturb the observer
dynamics, rather than sensor measurements and/or plant dynamics. 
\hfill$\Box$
\end{remark}

Since the aim is to obtain an algorithm for fending such biasing attacks,
we assume that the matrices $F_i$ are known to the defender. That is, the
observer (\ref{UP7.C.d}) is regarded as a model of the misappropriated
attack perceived by the defender. The defender has several choices for the
matrices $F_i$. For instance, one can assume 
$F_i=[1,\ldots,1]'$; this choice captures the attack model where the
attacker injects a scalar biasing input into the observer
dynamics. Alternatively, we can consider $F_i=I$; this means that the
adversary may bias each component of the observer separately. 

We assume that the attacker does not seek to change the network
topology. Although it is conceivable that the adversary may  
attempt to do so, this type of attack will imminently expose the
attacker. It may not be suitable for the adversary who wishes to remain
covert.

\begin{definition}[\cite{DUSL1}]\label{def.admiss}
A class of admissible biasing inputs, denoted
$\mathcal{F}_a$, consists of causal signals $f(t)\in \mathbf{R}^{n_f}$,
\begin{equation}
  \label{decomp}
f(t)=f_{1}(t)+f_{2}(t),  
\end{equation}
where the Laplace transform of $f_{1}(t)$, $f_{1}(s)$, has the property
$f_1^\infty=\sup_{\omega}|\omega f_{1}(j\omega)|^2<\infty$,  and
$f_{2}\in L_2[0,\infty)$. In the decomposition (\ref{decomp}), $f_{1}(t)$
represents the biasing component, and 
$f_{2}(t)$ is a masking signal; cf.~\cite{TSSJ-2015}.  
\end{definition}

   Note that the attack set  $\mathcal{F}_a$ includes as a special case
biasing attack inputs which consist of an
unknown steady-state component and an exponentially vanishing masking
signal generated by a low pass filter~\cite{TSSJ-2015}. 

The following lemma characterizes the properties of admissible biasing
inputs. Its proof is given in~\cite{DUSL1}. 

\begin{lemma}\label{admiss.f} Let $N(s)$, $D(s)$ be arbitrary real polynomials
  with the following properties: 
  \begin{enumerate}[(a)]
  \item
    The degree of $N(s)$ is not greater than the degree of $D(s)$, and so
    the transfer function $G(s)=\frac{N(s)}{D(s)}I$ is proper; 
  \item
    The transfer function $(sI+G(s))^{-1}G(s)$
    is stable, and hence 
$g_1\triangleq\frac{1}{2\pi}\int_{-\infty}^{\infty}\left|\frac{D(j\omega)}{j\omega 
    D(j\omega)+N(j\omega)}\right|^2d\omega<\infty$ and $g_2\triangleq \sup_\omega \left|\frac{j\omega
    D(j\omega)}{j\omega D(j\omega)+N(j\omega)}\right|^2
<\infty$.
  \end{enumerate}
Then for any signal $f\in \mathcal{F}_a$, it holds that $\nu\triangleq \hat
f-f\in L_2[0,\infty)$, where $\hat f$ is the signal whose Laplace transform is
  \begin{equation}
    \label{eq:2}
    \hat f(s)=\frac{1}{s}(sI+G(s))^{-1}G(s)f(s).
  \end{equation}
\end{lemma}

The idea behind the decomposition (\ref{decomp}) is to separate `slow'
parts of $f(t)$ responsible for biasing the observer (denoted
  $f_1(t)$) from disturbance-like 
components (denoted
  $f_2(t)$) whose impact can be attenuated provided the observer is
made sufficiently robust to disturbances. According to Lemma~\ref{admiss.f}, 
sufficiently slow biasing inputs can be approximated with a dynamic model, up
to a bounded energy error. This dynamic model will be used in the
derivation of a distributed observer. The idea is to endow the observer
with a capacity to filter out slow biases and be robust against 
bounded-energy perturbations and attack approximation errors alike.  

\begin{remark}\label{Comment2-1}
In this paper, the attacker is assumed to have no knowledge of the system, the
observers or the network. This forces the adversary to structure the
biasing signal to include both slow-varying biasing components and
disturbance-like components 
into the signal $f(t)$. If the slow 
biasing component is not included in $f(t)$, and $f(t)$ acts as a
disturbance, the observer will likely attenuate its effect along with the
effect of other noises and disturbances present in the system, since it is
designed 
to be robust against disturbances. This will likely reduce effectiveness of
the  attack. 
This motivates our assumption that the admissible biasing inputs must have
the form (\ref{decomp}). 

In contrast, the Byzantine model does
not prescribe attack inputs to have a particular structure. However, within
the Byzantine model, 
the attacker is assumed to have  
knowledge of the system which is not required in our model. For instance 
in~\cite{MS-2016}, the adversarial nodes are assumed   
``to possess complete
knowledge about the graph topology and the plant dynamics, i.e., an
adversarial node knows the measurements received by the normal nodes at
every time step.'' The attack model (\ref{decomp}) does not use this
assumption. Also, the Byzantine model limits the number of adversarial
nodes that can be tolerated. If this number exceeds a certain threshold,
this model cannot guarantee a successful attack detection. In contrast, our
model does not impose such a threshold, it allows for the situation when all
nodes are corrupted with biasing 
inputs. This indicates that both models have place in the theory of
resilient estimation, with their advantages and limitations. 
\hfill$\Box$
\end{remark}

\subsection{Resilient distributed estimation problem}\label{Problem.formulation}

Our proposal is to modify the observers (\ref{UP7.C.d.unbiased}) to include
additional control inputs to suppress biasing misappropriation attacks. The
corresponding model of a misappropriated observer will then be as follows
\begin{eqnarray}  
    \dot{\hat x}_i&=&A(t)\hat x_i + L_i^r(t)(y_i(t)-C_i(t)\hat x_i)-F_iu_i \nonumber \\
&& +\sum_{j\in
      \mathbf{V}_i}K_{ij}^r(t)(c_{ij}-W_{ij}\hat x_i)+F_if_i,
  \label{UP7.C.d.res} \quad 
    \hat x_i(0)=\xi_i.\quad 
\end{eqnarray}
The gains $L_i^r(t)$, $K_{ij}^r(t)$ of
the observer will be computed to ensure that each $\hat x_i$
converges to $x(t)$ even in the presence of admissible attacks.
To achieve this resilience property, the control inputs $u_i$ will need to be 
generated so that when node $i$ is under attack, $u_i$ counters the
biasing input $f_i$. Also, when the node $i$ is not attacked directly,
the control $u_i$ must not impede its observer from producing unbiased
estimates of $x(t)$. Since the signals $\hat f_i$ approximate $f_i$ up to an
$L_2$-integrable error $\nu_i$, this will be achieved by
forcing  $u_i$ to track $\hat f_i$ instead of $f_i$ while attenuating the
errors $\nu_i$ along with the system disturbances  $w$, $v_i$,
$v_{ij}$.   
 
To generate suitable control inputs, 
each node observer will be augmented with an output feedback
controller  
\begin{eqnarray}
  \dot\chi_i&=&\mathcal{A}_c(t)\chi_i
+\mathcal{B}_{c,i}(\zeta_i,\zeta_{ij},\eta_{ij},j\in\mathbf{V}_i) +F_{c,i}f_i-F_{c,i}u_i, \nonumber 
\\
  u_i&=&C_{c,i}(t)\chi_i, \qquad
  \chi_i(0)=\chi_{i,0}.   \label{detector.general}
\end{eqnarray}

The second last term in (\ref{detector.general}) captures the
situation where the attacker interferes with the operation
of both the state observer \emph{and} the defence layer at the
misappropriated node, and the last term is included to compensate this
interference, in the same manner as this is done for the main
observer. The matrices $F_{c,i}$ are selected by the   
defender. Similar to the matrices $F_i$, they describe the
attack pattern anticipated by the defender.  

The inputs to the controller (\ref{detector.general}) are the innovation
signals  
\begin{eqnarray}
  \zeta_i&=&y_i-C_i(t)\hat x_i, \quad \zeta_{ij}=c_{ij}-W_{ij}\hat x_i, \label{out.y}  \label{out.c}
\end{eqnarray}
which capture the new information contained in the local measurements
and obtained through communications, respectively. These signals are
readily available at node $i$ and will be used for both 
estimating the plant and detecting and compensating biasing attacks. Also,
controllers  (\ref{detector.general}) will be allowed to communicate, and 
the signals
\begin{eqnarray}
  \eta_{ij}&=&W_{c,ij}\chi_j+H_{c,ij}v_{c,ij},
                                              \label{out.chi}
\end{eqnarray}
describe the information received from the neighbors' controllers 
through imperfect channels containing disturbances $v_{c,ij}\in L_2[0,\infty)$. 
For simplicity, we assume that these communications replicate the
topology of the original network $\mathbf{G}$,  
because in practice the same channels will likely be used to transmit both
$c_{ij}$ and $\eta_{ij}$. 

The problem of resilient estimation is now formally stated. Let
$e_i(t)=x(t)-\hat x_i(t)$ be the state estimation error of the observer
(\ref{UP7.C.d.res}) at node $i$, and define the vector of observer
errors $\mathbf{e}=[e_1'~\ldots~e_N']'$. 

\begin{definition}\label{Problem1}
The problem of resilient estimation in this paper is concerned with constructing
a network of controlled observers (\ref{UP7.C.d.res}),
(\ref{detector.general}), to achieve the following: 
\begin{enumerate}[(i)]
\item
In the absence of disturbances and when the system is not under attack,
$e_i$ and $u_i$ must converge to 0 exponentially as $t\to\infty$  at
every node $i$.  
\item
In the presence of uncertainties and/or when the system is subjected to an
attack of class $\mathcal{F}_a$, the network of controlled observers
(\ref{UP7.C.d.res}),  (\ref{detector.general}) must ensure that 
$\int_0^{\infty}\|f_i-u_i\|^2dt<\infty$ $\forall i=1, \ldots, N$, and also
$\int_0^{\infty}\mathbf{e}'P\mathbf{e}dt < \infty$, 
Here $P=P'\ge 0$ is a given 
$nN\times nN$ matrix. 
\end{enumerate}
\end{definition}

We will subsequently show that when $f_i=0$ and the node $i$ is not
under attack, then the signals $u_i(t)$ vanish asymptotically even in the
presence of disturbances. Thus, the asymptotic behavior of the control
signals $u_i$ 
at the compromised nodes differs from the behavior of similar signals at
healthy nodes, which makes the signals $u_i$ suitable to be used as the attack
indicators. Also according to (ii), in the event of attack, the estimates
produced by the network of controlled observers  (\ref{UP7.C.d.res}), (\ref{detector.general}) will remain unbiased, up to
an $L_2$ integrable weighted error $P^{1/2}\mathbf{e}$. We will subsequently
guarantee a certain level of disturbance attenuation
with respect to this weighted error.   

\section{Design of a resilient distributed observer}\label{resilient.estimation}
 
\subsection{Observer error dynamics}

Consider the
dynamics of the estimation errors of the controlled observers
(\ref{UP7.C.d.res}),
 \begin{eqnarray}
    \dot{e}_i&=&(A(t) - L_i^r(t)C_i(t)-\sum_{j\in
      \mathbf{V}_i}K_{ij}^r(t)W_{ij})e_i + F_iu_i\nonumber \\ &+&\!\!\sum_{j\in
      \mathbf{V}_i}K_{ij}^r(t)W_{ij}e_j +B(t)w-L_i^r(t)D_i(t)v_i  \nonumber \\ & 
-& \!\!\sum_{j\in
      \mathbf{V}_i}K_{ij}^r(t)H_{ij}v_{ij} 
      -F_if_i, \label{e} 
\quad e_i(0)=x_0-\xi_i. \quad
\end{eqnarray}

Also, for every node $i$, consider a class of admissible biasing inputs
$\mathcal{F}_a$, of dimension 
$n_{f_i}$. According to Lemma~\ref{admiss.f}, a proper $n_{f_i}\times
n_{f_i}$ transfer function $G_i(s)$ can be associated
with each class of admissible biasing inputs. Let $\hat f_i(t)$, $\nu_i(t)$
denote the corresponding approximation signal defined by
(\ref{eq:2}) and the corresponding approximation error $\nu_i(t)=\hat
f_i(t)-f_i(t)$. From (\ref{eq:2}), $\nu_i$ and  $\hat f_i$ are
related as $\hat f_i=-\frac{1}{s}G_i(s)\nu_i$. Consider the minimal
realization of the transfer function $-\frac{1}{s}G_i(s)$:
\begin{eqnarray}
&&\dot\epsilon_i = \Omega_i\epsilon_i+\Gamma_i \nu_i, \qquad
\epsilon_i(0)=0, \label{Om.sys.general} \\ 
&&\hat f_i= \Upsilon_i\epsilon_i, \nonumber
\end{eqnarray}
Next, let us substitute $f_i=\Upsilon_i\epsilon_i-\nu_i$ into
equation (\ref{e}) and combine the dynamics of
the systems (\ref{e}) and (\ref{Om.sys.general}) into a system
with $(e_i',\epsilon_i')'$ as a state vector, 
\begin{eqnarray}
    \dot{e}_i&=&(A(t) - L_i^r(t)C_i(t)-\sum_{j\in
      \mathbf{V}_i}K_{ij}^r(t)W_{ij})e_i \nonumber \\ &+&\sum_{j\in
      \mathbf{V}_i}K_{ij}^r(t)W_{ij}e_j -F_i\Upsilon_i\epsilon_i+F_iu_i
    \nonumber \\ &+& 
B(t)w-L_i^r(t)D_i(t)v_i  
-\sum_{j\in
      \mathbf{V}_i}K_{ij}^r(t)H_{ij}v_{ij} 
      +F_i\nu_i,  \nonumber \\
\dot\epsilon_i &=& \Omega_i\epsilon_i+\Gamma_i \nu_i, \label{ext.e} 
\qquad e_i(0)=x_0-\xi_i, \quad \epsilon_i(0)=0.  
\end{eqnarray}
The system (\ref{ext.e}) is driven by
the bounded energy disturbances $w$, $v_i$, $v_{ij}$ and the error signals
$\nu_i$ which also have bounded energy, according to Lemma~\ref{admiss.f}.
Each such system is controlled via the control input $u_i$.
The innovation signals
(\ref{out.y}) can be written in terms of the variables of
the systems (\ref{ext.e}) as 
\begin{eqnarray}
  \zeta_i&=&C_i(t)e_i + D_iv_i, \quad 
  \zeta_{ij}=-W_{ij}(e_j-e_i)+H_{ij}v_{ij}. \quad
                                          \label{out.y.1} \label{out.c.1}
\end{eqnarray}
Therefore, they can be regarded as outputs of the interconnected large-scale
uncertain system comprised of systems (\ref{ext.e}). The attack  tracking
error  $f_i-u_i$ can also be expressed in terms of this system variables,
\begin{equation}
  \label{eq:3}
  \varphi_i=f_i-u_i=\Upsilon_i\epsilon_i-u_i-\nu_i.
\end{equation} 
It will be regarded as the system performance output.
This allows to carry out the derivation of the observers (\ref{UP7.C.d.res}) and
controllers of the form  (\ref{detector.general}) within the 
$H_\infty$ disturbance attenuation framework. This approach will be pursued
in the remainder of the paper.    

\subsection{Feedback attack detector and compensator}

Since  the biasing inputs $f_i$ 
are not available directly, we first introduce a distributed filter
to estimate the state of the extended system (\ref{ext.e}) from the
information available at the nodes, namely from 
the outputs (\ref{out.y}) and interconnections $\eta_{ij}$:
\begin{eqnarray}
    \dot{\hat{e}}_i&=&(A(t) - L_i^r(t)C_i(t)-\sum_{j\in
      \mathbf{V}_i}K_{ij}^r(t)W_{ij})\hat{e}_i \nonumber \\ 
& +& \sum_{j\in
      \mathbf{V}_i}K_{ij}^r(t)\eta_{ij} 
+\bar L_i^r(\zeta_i-C_i(t)\hat{e}_i)
    \nonumber \\ 
    &+& 
\sum_{j\in
      \mathbf{V}_i}\bar K_{ij}^r(t)(\zeta_{ij}-W_{ij}\hat{e}_i+\eta_{ij})
    + \bar F_if_i-\bar F_iu_i,
\nonumber \\
  \dot{\hat\epsilon}_i &=& 
\Omega_i \hat\epsilon_i +
    \check L_i^r(t)(\zeta_i-C_i(t)\hat{e}_i) \nonumber \\
&+&\sum_{j\in
      \mathbf{V}_i}\check
    K_{ij}^r(t)(\zeta_{ij}-W_{ij}\hat{e}_i+\eta_{ij})
+\check F_if_i-\check F_iu_i, \quad
     \label{ext.obs.nu.1.om.res}  \\
&&\hat{e}_i(0)=0, \quad \hat\epsilon_i(0)=0. 
\nonumber
\end{eqnarray}
Each filter (\ref{ext.obs.nu.1.om.res}) has the desired form of the system
(\ref{detector.general}); i.e., $\chi_i=[\hat e_i',\hat\epsilon_i']'$. To
complete the similarity with 
(\ref{detector.general}), define the interconnections (\ref{out.chi})
between the filters  (\ref{ext.obs.nu.1.om.res}) and their outputs $u_i$ as
follows:  
\begin{eqnarray}
  \label{eq:eta}
\eta_{ij}&=&W_{ij}\hat e_j+H_{c,ij}v_{c,ij}, \\
  \label{varphi}
  u_i&=& \Upsilon_i\hat{\epsilon}_i;   
\end{eqnarray}
i.e., $W_{c,ij}=[W_{ij}~0]$, $C_{c,i}=[0~\Upsilon_i]$. 

The distributed filter comprised of the systems
(\ref{ext.obs.nu.1.om.res})--(\ref{varphi}) will be used to generate 
control inputs for the underlying observer (\ref{UP7.C.d.res}).  
As explained previously, it accounts for the possibility that the adversary may
launch a biasing attack on the controller, along with attacking  
the original observer. Also, the feedback is included to compensate for
these attacks.

\begin{remark}
The use of feedback for compensating biasing attack inputs is the main
distinction between the systems (\ref{ext.obs.nu.1.om.res})--(\ref{varphi})
and similar filters
proposed in~\cite{DUSL1,U10a} for detecting biasing attacks. 
Thus, the role of the systems (\ref{ext.obs.nu.1.om.res})--(\ref{varphi})
in this paper shifts from signaling biasing attacks to countering them.
\end{remark}

\subsection{The system design procedure}


Consider the errors of the filters (\ref{ext.obs.nu.1.om.res}),
$z_i=e_i-\hat e_i$, $\delta_i=\epsilon_i-\hat\epsilon_i$. It follows
from (\ref{ext.e}) and (\ref{eq:3})--(\ref{varphi}) that 
\begin{eqnarray}
    \dot{z}_i&=&(A(t) - \hat L_i^r(t)C_i(t)-\sum_{j\in
      \mathbf{V}_i}\hat K_{ij}^r(t)W_{ij})z_i - \hat F_i\Upsilon_i\delta_i + Bw 
    \nonumber \\ &+& \sum_{j\in
      \mathbf{V}_i}\hat K_{ij}^r(t)(W_{ij}z_j-\hat H_{ij}\hat v_{ij}) -\hat L_i^r(t)D_i(t)v_i 
      +\hat F_i\nu_i, \nonumber \\
\dot \delta_i&=&\Omega_i \delta_i-\check L_i^r(t)C_i(t)z_i-\sum_{j\in
      \mathbf{V}_i}\check K_{ij}^r(t)W_{ij}z_i - \check F_i\Upsilon_i\delta_i
    \nonumber \\
&+& \sum_{j\in
   \mathbf{V}_i}\check K_{ij}^r(t)(W_{ij}z_j- \hat H_{ij}\hat v_{ij})  -\hat L_i^r(t)D_i(t)v_i + \check
    \Gamma_i \nu_i, \nonumber 
\\
&&  z_i(0)=x_0-\xi_i, \quad \delta_i(0)=0. \label{ext.error.0} \label{ext.error}
\end{eqnarray}
Here we have combined all channel disturbances into the vectors $\hat
v_{ij}=[v_{ij}'~v_{c, ij}']'$ and 
have also  used the following notation: 
\begin{eqnarray}
&&\hat L_i^r(t)=L_i^r(t)+\bar L_i^r(t), \quad 
\hat K_{ij}^r(t)=K_{ij}^r(t)+\bar K_{ij}^r(t), \nonumber \\
&& \hat F_i=F_i+\bar F_i, \quad \check \Gamma_i=\Gamma_i+\check F_i, 
\quad \hat H_{ij}=[H_{ij}~H_{c,ij}]. \qquad  
\label{LK}
\end{eqnarray}
The signal $\varphi_i$ defined in (\ref{eq:3}) will serve as a
performance output for the system (\ref{ext.error}), since 
$  \varphi_i=\Upsilon_i\delta_i-\nu_i$. 

This notation reveals that the disturbance attenuation properties of
the system comprised of systems (\ref{ext.error}) can be analyzed
separately from the underlying resilient estimation problem. This
leads us to propose the procedure for obtaining the coefficients
for the observers (\ref{UP7.C.d.res}) and the controllers
(\ref{ext.obs.nu.1.om.res})--(\ref{varphi}):

\begin{enumerate}[1.]
\item
First, the coefficients $\hat L_i^r(t)$, $\hat K_{ij}^r(t)$, $\check
L_i^r(t)$, $\check K_{ij}^r(t)$ for each system (\ref{ext.error})  
will be derived. The aim of this derivation is to guarantee that the filters
(\ref{ext.obs.nu.1.om.res})--(\ref{varphi}) are able to track the signals
$\hat f_i$. The following specific properties are sought: 
\begin{enumerate}[(i)]
\item
  When $w=0$, $v_i=0$, and $\hat v_{ij}=0$, $j\in\mathbf{V}_i$,
  $i=1,\ldots, N$, and the system is not under attack, i.e., when $f_i=0$ for
  all $i=1, \ldots,N$, all signals $z_i$, $\delta_i$ must decay to 0
  exponentially. 
\item
  When at least one of the signals $w$, $v_i$, $\hat v_{ij}$ or $f_i$ is
  nonzero, all signals $z_i$, $\delta_i$ must vanish asymptotically,  
and the following disturbance attenuation property must hold 
\begin{eqnarray}
\lefteqn{\hspace{-5ex}\sum_{i=1}^N\int_0^{\infty}\|\Upsilon_i\delta_i\|^2dt\le \gamma^2 \sum_{i=1}^N\bigg(\|x_0-\xi_i\|_{X_i^{-1}}^2} &&
                                                        \nonumber \\
&&\hspace{-5ex}+ \int_0^\infty\!\!\big(\|w\|^2+\|v_i\|^2+\|\nu_i\|^2+\sum_{j\in
    \mathbf{V}_{ij}}\|\hat v_{ij}\|^2\big)dt\bigg). \quad
\label{delta.Hinf.i}\label{z.Hinf}
\end{eqnarray}  
Here $\gamma^2$ denotes the level of disturbance attenuation which
reflects a desired tracking accuracy.
\end{enumerate}

\item
Next, the coefficients $L_i^r(t)$, $K_{ij}^r(t)$ for the controlled distributed
observer (\ref{UP7.C.d.res}) will be computed. With the parameters 
$\hat L_i^r(t)$, $\hat K_{ij}^r(t)$, $\check
L_i^r(t)$, $\check K_{ij}^r(t)$ obtained
in the previous step, it follows from (\ref{delta.Hinf.i}) that the
tracking error $\varphi_i=\Upsilon_i\delta_i-\nu_i$ is  
$L_2$ integrable for every attack input $f_i\in\mathcal{F}_a$,
therefore the coefficients 
$L_i^r(t)$, $K_{ij}^r(t)$ will be obtained so that the observer
(\ref{UP7.C.d.res}) attenuates this additional disturbance, along with $w$,
$v_i$, $\hat v_{ij}$. Essentially, we redesign the original unbiased  
observers (\ref{UP7.C.d.unbiased}) to endow them with robustness 
against attack tracking errors. This additional robustness property is
needed to ensure the observers can withstand the transients generated by 
the attack canceling controllers.

\item
Finally, the remaining coefficients $\bar L_i^r(t)$, $\bar K_{ij}^r(t)$ of
the attack detector (\ref{ext.obs.nu.1.om.res}) will be obtained from
(\ref{LK}), using the values of $\hat L_i^r(t)$, $\hat K_{ij}^r(t)$ and
$L_i^r(t)$, $K_{ij}^r(t)$ obtained at the previous steps.   
\end{enumerate}


To implement this procedure, we employ a decentralized
observer synthesis technique from~\cite{ZU1a}. It has an advantage in that
it the parameters of the nodes 
observers are computed in a decentralized fashion. This feature is particularly
attractive in the presence of adversaries, since each node observer can be 
computed on-line independently from other nodes, after a certain initial
centralized setup. 

The following technical result is adapted from~\cite{ZU1a}. Consider an
interconnected system comprised of the systems 
 \begin{eqnarray}
    \dot{\tilde{e}}_i&=&(\tilde A_i(t) - \tilde L_i(t)\tilde C_i(t)-\sum_{j\in
      \mathbf{V}_i}\tilde K_{ij}(t)\tilde W_{ij})\tilde e_i \nonumber \\ &+&\!\!\sum_{j\in
      \mathbf{V}_i}\tilde K_{ij}(t)\tilde W_{ij}\tilde e_j +\tilde
    B_i(t)\tilde w_i-\tilde L_i(t)D_i(t)v_i  \nonumber \\ & 
-& \!\!\sum_{j\in
      \mathbf{V}_i}\tilde K_{ij}(t)\tilde H_{ij}\tilde v_{ij} \label{t.e} , \quad \tilde e_i(0)=\tilde e_{0,i}, \nonumber
\end{eqnarray}
subject to $L_2$-integrable disturbances $\tilde w_i$, $v_i$, $\tilde v_{ij}$. The
matrix-valued functions  $\tilde A_i$, $\tilde C_i$, and $\tilde B_i$ are
 bounded on $[0,\infty)$.  
 
Let $\tilde X_i$, $\tilde R_i$,
be positive definite symmetric $n\times n$ matrices, and $\tilde Z_{ij}$, $j\in
\mathbf{V}_i$, $i=1,\ldots, N$, be positive definite symmetric
$p_{ij}\times p_{ij}$ matrices. Also, define the following matrices: $\tilde U_{ij}=\tilde H_{ij}\tilde H_{ij}'+\tilde Z_{ij}>0$,
\begin{eqnarray}
\tilde \Delta_i&=&\sum_{j\in\mathbf{V}_i}\tilde W_{ij}'\tilde
U_{ij}^{-1}\tilde Z_{ij}\tilde U_{ij}^{-1}\tilde W_{ij}, 
\label{t.Phi.def.1} \\
\tilde \Phi_{ij}&=&\begin{cases}\tilde \Delta_i, & i=j,\\
-\tilde W_{ij}'\tilde U_{ij}^{-1}\tilde W_{ij}, & i\neq j,~j\in\mathbf{V}_i,\\
0 & i\neq j,~j\not\in\mathbf{V}_i.
\end{cases}
\label{t.Phi.def} 
\end{eqnarray}
Let
$\tilde R\triangleq \mbox{diag}[\tilde R_1,\ldots, \tilde R_N]$, $\tilde \Delta\triangleq
\mbox{diag}[\tilde \Delta_1, \ldots, \tilde \Delta_N]$ and $\tilde \Phi\triangleq
[\tilde \Phi_{ij}]_{i,j=1,\ldots, N}$. Next, 
consider $N$ differential Riccati equations: 
\begin{eqnarray}
  \label{ARE}
  \dot {\tilde Y}_i&=& \tilde A_i\tilde Y_i+\tilde Y_iA'+ \tilde B_i \tilde
B_i'-\tilde Y_i\Big(
\tilde C_i(t)'(D_i(t)D_i(t)')^{-1}\tilde C_i(t) \nonumber \\
&+&\sum_{j\in\mathbf{V}_i} \tilde W_{ij}'\tilde U_{ij}^{-1}\tilde W_{ij}
-\frac{1}{\tilde \gamma^2}\tilde R_i\Big)\tilde Y_i, \quad 
\tilde Y_i(0)=\tilde X_i^{-1}. \quad 
\end{eqnarray}
Each equation (\ref{ARE}) only depends on the parameters associated with
node $i$, and can be solved by node $i$ without
interacting with its neighbors. 

\begin{lemma}[cf.~\cite{ZU1a}]\label{stabt}
For a given a positive semidefinite $nN\times nN$ matrix
$\tilde P=\tilde P'$ and a constant $\tilde \gamma>0$, suppose the collection of matrices
$\tilde X_i$, $\tilde R_i$ and $\tilde Z_{ij}$,
$j\in\mathbf{V}_i$, $i=1,\ldots, N$,
is found which satisfies the conditions:
\begin{enumerate}[(a)]
\item
  It holds that  
\begin{eqnarray}\label{Minv}
\tilde R > \tilde P-\tilde \gamma^2(\tilde \Phi+\tilde \Phi'-\tilde \Delta), 
\end{eqnarray}

\item
The equations~(\ref{ARE}) have 
positive definite bounded 
solutions on $[0,\infty)$; i.e.,  $\alpha_{1i}I\le
\tilde Y_i(t)\le \alpha_{2i}I$ ($\exists \alpha_{1i},\alpha_{2i}>0$).
\end{enumerate}
Then, the interconnected system comprised of systems (\ref{t.e})
equipped  with the coefficients 
\begin{eqnarray}
\tilde L_i(t)&=&\tilde Y_i(t) \tilde C_i(t)'(D_i(t)D_i(t)')^{-1}, \nonumber \\
\tilde K_{ij}(t)&=&\tilde Y_i(t) \tilde W_{ij}'\tilde U_{ij}^{-1},\qquad
  \label{Li}\label{Kij}
\end{eqnarray}
has the following properties:
\begin{enumerate}[(i)]
\item 
In the absence of disturbances, $\tilde e_i$ vanish  exponentially. 

\item
In the presence of disturbances, each $\tilde e_i(t)$ vanishes asymptotically,
and it holds that 
\begin{eqnarray}
  \label{eq:5}
\lefteqn{\int_0^{\infty}\tilde e '\tilde P{\tilde e}dt\le \tilde \gamma^2
  \sum_{i=1}^N\bigg(\|\tilde e_{0,i}\|_{\tilde X_i}^2}
&& 
\nonumber \\
&&+ \int_0^\infty\big(\|\tilde w_i\|^2+\|v_i\|^2+\sum_{j\in
    \mathbf{V}_{ij}}\|\tilde v_{ij}\|^2\big)dt \bigg). \quad
\end{eqnarray}
\item
In addition, each system (\ref{t.e}) with the coefficients defined in   
(\ref{Kij}) attenuates the local disturbances $\tilde w_i(t)$, $v_i$, 
$\tilde v_{ij}$ and its neighbors inputs $\tilde e_j$,
$j\in\mathbf{V}_i$: 
\begin{eqnarray}
\lefteqn{\int_0^\infty \tilde e_i'(t) \tilde R_i \tilde e_i(t)dt} && \nonumber \\
&\le& \tilde \gamma^2 
\bigg(\|\tilde e_{0,i}\|^2_{\tilde X_i}+\int_0^\infty\Big(\|\tilde w_i(t)\|^2+
  \|v_i(t)\|^2  \nonumber \\
    && +\sum_{j\in
    \mathbf{V}_i}(\|\tilde v_{ij}(t)\|^2+ \|\tilde e_j(t)\|_{Z_{ij}^{-1}}^2)
\Big)dt\bigg).
\label{eq:8}
\end{eqnarray}
\end{enumerate}
\end{lemma}

The proof of the lemma is similar to the proof of Theorem~1 in~\cite{ZU1a};
it is omitted for brevity.  

  \begin{remark}\label{Comment2-7}
In Lemma~\ref{stabt}, each node
exercises its own $H_\infty$ filter, and uses its own
differential Riccati equation~(\ref{ARE}) to compute the filter.
The existence of this filter is equivalent to the 
Riccati equation~(\ref{ARE}) having a bounded solution. A standard proof of
this fact~\cite{NK91} uses a certain detectability
assumption. In our case, it requires the pair  $(\tilde A(t), [\tilde C_i(t)', \tilde
  W_{i1}', \ldots, \tilde W_{ij_{d_i}}']')$ to be detectable; $d_i$ is the
  cardinality of $\mathbf{V}_i$.  
However, this detectability property is only needed to prove the
necessity of the Riccati equation condition. Since
the conditions of Lemma~\ref{stabt} are sufficient, the above detectability
assumption does not need to be stated explicitly; it is subsumed in the
requirement of the lemma that the Riccati equation (\ref{ARE}) must have a
bounded solution.  When this true, each local
node error dynamics system is dissipative with the storage function
$\tilde V_i=\tilde e_i'\tilde Y_i^{-1} \tilde e_i$ and the supply rate
$
\tilde \gamma^2 \left(\|\tilde w_i\|^2+\|v_i\|^2+\sum_{j\in\mathbf{V}_i}(\|\tilde
v_{ij}\|^2 + \|\tilde e_j\|^2_{Z_{ij}^{-1}})\right)-\tilde e_i'\tilde R_i
\tilde e_i.  
$
The inequality (\ref{Minv}) balances these dissipativity characteristics of
individual nodes in such a way that the 
error dynamics of the whole network are dissipative, with the storage function
$\tilde V=\sum_{i=1}^N\tilde e_i'\tilde Y_i^{-1} \tilde e_i$ and the supply
rate  
$
\tilde \gamma^2 \sum_{i=1}^N(\tilde w_i\|^2+\|v_i\|^2+\sum_{j\in\mathbf{V}_i}\|\tilde
v_{ij}\|^2)-\tilde e'\tilde P \tilde e; 
$
also, see~\cite{ZU1a}.  
\end{remark} 

Lemma~\ref{stabt} will play an instrumental role in the 
procedure for the design of resilient filters proposed in this paper. We
are now in a position to describe this procedure in detail. 

\subsubsection*{Step 1. Stabilization of the detector error dynamics
  (\ref{ext.error}) via output injection}    
Introduce the following notation:
\begin{eqnarray}
\mathbf{A}_i(t)&=&\left[\begin{array}{cc}
A(t) & -F_i\Upsilon_i\\
0 & \Omega_i
    \end{array}
  \right], \quad 
\mathbf{B}_i=\left[\begin{array}{cc}
B(t) & \hat F_i\\
0 & \check \Gamma_i
    \end{array}
  \right], \nonumber \\
\mathbf{C}_i(t)&=&\left[\begin{array}{cc}
C_i(t) & 0\end{array}\right], \quad
 \mathbf{W}_{ij}(t)=\left[\begin{array}{cc}                                  
         W_{ij} & 0 \end{array}\right], \\
\mathbf{L}_i^r&=&\left[\begin{array}{c}
\hat L_i^r \\
\check L_i^r\end{array}\right], \quad
  \mathbf{K}_{ij}^r=\left[\begin{array}{c}
 \hat K_{ij}^r  \\                          
\check K_{ij}^r
\end{array}\right]. \quad                                               
                                            \label{Ldef} 
\end{eqnarray}
Also, consider  positive definite
$(n+n_{f_i})\times (n+n_{f_i})$ 
block-diagonal matrices 
$\mathbf{X}_i$, $i=1\ldots,N$, partitioned as
$\mathbf{X}_i=\left[\begin{array}{cc}X_i & 0\\ 0 & X_{i,0}
  \end{array}\right]$,
where $X_i=X_i'>0$, $X_{i,0}'=X_{i,0}>0$ are respectively $n\times n$ and
$n_{f_i}\times n_{f_i}$ matrices. In addition, consider the matrices $\Phi$ and $\Delta$
of the form (\ref{t.Phi.def}), (\ref{t.Phi.def.1}):
\begin{eqnarray}
\Delta_i&=&\sum_{j\in\mathbf{V}_i} W_{ij}'U_{ij}^{-1}Z_{ij}U_{ij}^{-1} W_{ij},
\label{Phi.def.1} \\
\Phi_{ij}&=&\begin{cases} \Delta_i, & i=j,\\
- W_{ij}'U_{ij}^{-1} W_{ij}, & i\neq j,~j\in\mathbf{V}_i,\\
0 & i\neq j,~j\not\in\mathbf{V}_i,
\end{cases}
\label{Phi.def} 
\end{eqnarray}
where $Z_{ij}'=Z_{ij}$ and $U_{ij}=\hat
H_{ij}\hat H_{ij}'+ Z_{ij}$. 

\begin{theorem}\label{T1}
Let a constant $\gamma>0$ and symmetric matrices $R_i> 0$, 
$\check R_i> 0$, $Z_{ij}>0$ of dimensions $n\times n$, $n_{f_i}\times
n_{f_i}$, $p_{ij}\times p_{ij}$, respectively, $j\in \mathbf{V}_i$, $i=1,
\ldots N$, be selected so that the following conditions hold:     
\begin{enumerate}[(a)]
\item 
It holds that 
\begin{equation}
\label{LMI}
R+\gamma^2(\Phi+\Phi'-\Delta) > 0, \quad 
\check
  R_i>\Upsilon_i'\Upsilon_i,
\end{equation}
where $R=\mathrm{diag}[R_1, \ldots, R_N]$. 
\item
Each differential Riccati equation   
\begin{eqnarray}\label{Riccati} 
&&\hspace{-1.2cm}\dot{\mathbf{Y}}_i = \mathbf{A}_i\mathbf{Y}_i+\mathbf{Y}_i\mathbf{A}_i'+\mathbf{B}_i\mathbf{B}_i'
 - \mathbf{Y}_i\Big(\mathbf{C}_i'(D_iD_i')^{-1}\mathbf{C}_i
\nonumber \\
&&\hspace{-.7cm}+\sum_{j\in\mathbf{V}_i}\mathbf{W}_{ij}'U_{ij}^{-1}\mathbf{W}_{ij}                     
-\frac{1}{\gamma^2}\mathbf{R}_i\Big)\mathbf{Y}_i, \quad
\mathbf{Y}_i(0)=\mathbf{X}_i^{-1},  
\end{eqnarray}
with $\mathbf{R}_i\triangleq\left[\begin{array}{cc}R_i & 0\\ 0 & \check R_i
\end{array}\right]$,
has a positive definite symmetric bounded solution $\mathbf{Y}_i(t)$ on the
interval $[0,\infty)$.  
\end{enumerate}
Then the system comprised of systems (\ref{ext.error})
 with the coefficients $\hat L_i^r$, $\hat K_{ij}^r$, $\check L_i^r$,
 $\check K_{ij}^r$, obtained by partitioning the matrices
\begin{eqnarray}
  \label{L}
  \mathbf{L}_i^r(t)&=&\mathbf{Y}_i(t)\mathbf{C}_i(t)'(D_i(t)D_i(t)')^{-1}(t),
\nonumber  \\
  \mathbf{K}_{ij}^r(t)&=&\mathbf{Y}_i(t)\mathbf{W}_i'U_{ij}^{-1}
\end{eqnarray}
according to (\ref{Ldef}), has the following properties:
\begin{enumerate}[(i)]
\item 
In the absence of disturbances and attacks on the controlled observers
(\ref{UP7.C.d.res}), (\ref{ext.obs.nu.1.om.res}), $z_i(t)$, $\delta_i(t)$
vanish exponentially as $t\to\infty$, for all $i=1,\ldots,N$. 
\item
When the network of controlled observers (\ref{UP7.C.d.res}),
(\ref{ext.obs.nu.1.om.res}) is subjected to disturbances and/or admissible
biasing attacks of class $\mathcal{F}_a$, all $z_i(t)$, $\delta_i(t)$
decay asymptotically to 0 as $t\to\infty$, and 
(\ref{delta.Hinf.i}) holds. 
\end{enumerate}
\end{theorem}

\emph{Proof: } 
Using the notation (\ref{Ldef}) and
letting $\mu_i=[z_i'~\delta_i']'$, $\mathbf{w}_i=[w' ~\nu_i']'$, the
system~(\ref{ext.error}) can be written as
 \begin{eqnarray}
    \dot{\mu}_i&=&(\mathbf{A}_i(t) - \mathbf{L}_i^r(t)\mathbf{C}_i(t)-\sum_{j\in
      \mathbf{V}_i}\mathbf{K}_{ij}^r(t)\mathbf{W}_{ij})\mu_i \nonumber \\ &+&\!\!\sum_{j\in
      \mathbf{V}_i}\mathbf{K}_{ij}^r(t)\mathbf{W}_{ij}\mu_j +\mathbf{B}_i(t)\mathbf{w}_i-\mathbf{L}_i^r(t)D_i(t)v_i  \nonumber \\ & 
-& \!\!\sum_{j\in
      \mathbf{V}_i}\mathbf{K}_{ij}(t)^r\hat H_{ij}\hat v_{ij}, \label{e.mu} 
\quad \mu_i(0)=\left[
            \begin{array}{cc}
              (x_0-\xi_i)' & 0'
            \end{array}
\right]'. \nonumber
\end{eqnarray}
This system is precisely of the form of the system (\ref{t.e}) considered in
Lemma~\ref{stabt}. Note that in the absence of disturbances and attacks,
$\nu_i=0$ and $\mathbf{w}_i=0$; see (\ref{eq:2}). Therefore, claims (i) and
(ii) of the theorem can be inferred from the corresponding claims of  
Lemma~\ref{stabt}. For this, we need to validate the conditions of
that lemma.

To show that condition (a) of
Lemma~\ref{stabt} is satisfied, consider the block-diagonal matrix $\tilde
P$ composed of $N$ 
diagonal $(n+n_{f_i})\times (n+n_{f_i})$ blocks
$\tilde P_i=\left[\begin{array}{cc}
  0 & 0 \\
  0 & \Upsilon_i'\Upsilon_i
\end{array}\right]$; i.e.,  $\tilde P=\mathrm{diag}[\tilde P_1,\ldots,
\tilde P_N]$. Let $\tilde W_{ij}=\mathbf{W}_{ij}$. Then from the
definition of $\mathbf{W}_{ij}$, the matrices $\tilde
\Delta_i$, $\tilde \Phi_{ij}$
in (\ref{t.Phi.def}), (\ref{t.Phi.def.1}) are 
$
\tilde \Delta_i=\left[\begin{array}{cc}\Delta_i & 0 \\ 0 & 0
      \end{array}\right]$, $ \tilde \Phi_{ij}=\left[\begin{array}{cc}
  \Phi_{ij} & 0 \\
  0 & 0
\end{array}\right]$. 
The satisfaction of condition (\ref{Minv}) readily follows
from (\ref{LMI}) if we let  $\tilde R_i=\mathbf{R}_i$. Condition (b) of
Lemma~\ref{stabt} trivially follows from condition (b) of the theorem if we
let $\tilde A_i=\mathbf{A}_i$, $\tilde C_i=\mathbf{C}_i$, $\tilde
X_i=\mathbf{X}_i$. 
\hfill $\Box$

According to Theorem~\ref{T1}, each node can compute the matrices
$\mathbf{L}_i^r$, $\mathbf{K}_{ij}^r$ and their components $\hat L_i^r$,
$\hat K_{ij}^r$, $\check L_i^r$, $\check K_{ij}^r$ on-line, by solving the
respective Riccati equations (\ref{Riccati}). This allows to
compute the coefficients of the observer (\ref{UP7.C.d.res}) and the
controller (\ref{ext.obs.nu.1.om.res}) in real time. The nodes do not need
to communicate to solve these Riccati equations. 

Even though statement (iii) of Lemma~\ref{stabt} is not
  used explicitly in the proof of Theorem~\ref{T1}, together with
  the second inequality (\ref{LMI}) $\check
  R_i>\Upsilon_i'\Upsilon_i$ it yields
\begin{eqnarray*}
\lefteqn{\int_0^\infty \|\Upsilon_i\delta_i(t)\|^2dt< \gamma^2 
\bigg(\|z_{0,i}\|^2_{X_i}+\|\delta_{0,i}\|^2_{X_{i,0}}} && \\
&+&\int_0^\infty\Big(\|w(t)\|^2+\|\nu_i(t)\|^2+
  \|v_i(t)\|^2  \nonumber \\
    && +\sum_{j\in
    \mathbf{V}_i}(\|\hat v_{ij}(t)\|^2+ \|z_j(t)\|_{Z_{ij}^{-1}}^2)
\Big)dt\bigg).
\end{eqnarray*}
This inequality characterizes the capacity of the attack detector at node
$i$ to attenuate disturbances as well as the impact
of the neighbours' errors on the detector $i$'s
accuracy. In addition, the first inequality (\ref{LMI}),
$R+\gamma^2(\Phi+\Phi'-\Delta) > 0$, ensures that the interconnections
between the detector nodes are balanced so that the overall system is able
to absorb and dissipate the impact of these errors.\label{Comment2-6.2}    

Also note that the first inequality (\ref{LMI}) only
involves constant parameters of the interconnections between the
systems~(\ref{ext.obs.nu.1.om.res}). It does not involve  parameters of the
plant and the sensors, while the second inequality in
  (\ref{LMI}) only involves the output matrices of the minimal realization
  of $-\frac{1}{s}G_i(s)$ in~(\ref{Om.sys.general}).\label{Comment2-6.1}
This allows to solve  
the inequality (\ref{LMI}) off-line in 
advance, and then use its solutions $R_i$,  $\check R_i$, $Z_{ij}$ in
(\ref{Riccati}). Even if some of
the sensors at node $i$ have failed, and the dimension of the local sensor
matrices $C_i$, $D_i$ has changed as a result of this failure, one can
continue using the same matrices $R_i$, $\check R_i$, $Z_{ij}$. The Riccati
equation (\ref{Riccati}) at that node will have to be updated to include
the modified $C_i$, $D_i$. However, the computation of the filter gains at
other nodes will not be affected. This is a significant advantage, in
comparison with some existing 
distributed estimation techniques which require that the entire network of
observers must be recomputed should one of the sensors have failed.
To use this advantage, the matrix inequality
(\ref{LMI}) must be solved centrally to assign each node with suitable
matrices $R_i$, $\check R_i$\footnote{An obvious solution to~(\ref{LMI}) is 
$R_i=(\gamma^2\lambda_{\max}(\Phi+\Phi'-\Delta)+\alpha)I$ and $\check
R_i=(\lambda_{\max}(\Upsilon_i'\Upsilon_i)+\alpha)I$, where $\alpha>0$ is a
constant and $\lambda_{\max}(\cdot)$ is the largest eigenvalue of a
matrix. However not every solution of (\ref{LMI}) is suitable. As noted
previously, the matrices $R_i$ and $\check R_i$ must also be selected so
that the differential Riccati equations (\ref{Riccati}) have bounded
solutions.}. This needs to be done only once at time
$t=0$; this step can be regarded as an initialization of the
algorithm. Once the matrices 
$Z_{ij}$ are selected, the inequality (\ref{LMI}) 
becomes a linear matrix  inequality with respect to $R_i$, $i=1, \ldots, N$
and $\gamma^2$. It can be solved numerically using the existing
software. This feature facilitates tuning the local and global performance
of the proposed resilient filter.

\subsubsection*{Step 2. Design of the resilient distributed observer
  (\ref{UP7.C.d.res})}

From (\ref{UP7.C.d.res}), it is clear that the signals $\varphi_i$
in (\ref{eq:3}) affect the accuracy of the proposed 
resilient observer (\ref{UP7.C.d.res}). However, under conditions of
Theorem~\ref{T1}, $\tilde\nu_i=\Upsilon_i\delta_i$ is
$L_2$-integrable. Also, $\nu_i$ is $L_2$-integrable, according to
Definition~\ref{def.admiss}. Therefore, the effect of $\varphi_i$ can be
attenuated, along with the effects of $L_2$-integrable
disturbances present in the system.   
The coefficients $L_i^r$, $K_{ij}^r$ of the observer (\ref{UP7.C.d.res})
which accomplish this task can also be computed using Lemma~\ref{stabt}. 
To derive these coefficients, let us re-write the error dynamics
(\ref{e}) of the observer (\ref{UP7.C.d.res}) as
 \begin{eqnarray}
    \dot{e}_i&=&(A(t) - L_i^r(t)C_i(t)-\sum_{j\in
      \mathbf{V}_i}K_{ij}^r(t)W_{ij})e_i \nonumber \\ &+&\!\!\sum_{j\in
      \mathbf{V}_i}K_{ij}^r(t)W_{ij}e_j +B_{1,i}(t)w_i-L_i^r(t)D_i(t)v_i  \nonumber \\ & 
-& \!\!\sum_{j\in
      \mathbf{V}_i}K_{ij}(t)^rH_{ij}v_{ij}, \quad  
\label{e.res} 
e_i(0)=x_0-\xi_i; 
\end{eqnarray}
here we used the notation $
B_{1,i}(t)=
\left[
  \begin{array}{ccc}
    B(t) & -F_i 
  \end{array}
\right]$, $w_i=\left[
  \begin{array}{ccc}
    w' & \varphi_i' 
  \end{array}
\right]'$.
Then Lemma~\ref{stabt} can be applied to the interconnection of systems
(\ref{e.res}). As before, consider symmetric matrices $\bar Z_{ij}>0$, $\bar U_{ij}=H_{ij}H_{ij}'+\bar Z_{ij}>0$, $j\in 
\mathbf{V}_i$, $i=1, \ldots N$, and define the 
matrices $\bar\Phi=\left[\bar\Phi_{ij}\right]_{i,j=1}^N$,
$\bar\Delta=\mathrm{diag}[\bar\Delta_1, \ldots, \bar\Delta_N]$
of the form (\ref{t.Phi.def}), (\ref{t.Phi.def.1}): 
\begin{eqnarray}
\bar\Delta_i&=&\sum_{j\in\mathbf{V}_i}W_{ij}'\bar U_{ij}^{-1}\bar
Z_{ij}\bar U_{ij}^{-1}W_{ij}, \label{bar.Phi.def.1} \\
\bar\Phi_{ij}&=&\begin{cases}\bar\Delta_i, & i=j,\\
-W_{ij}'\bar U_{ij}^{-1}W_{ij}, & i\neq j,~j\in\mathbf{V}_i,\\
0 & i\neq j,~j\not\in\mathbf{V}_i.
\end{cases}
\label{bar.Phi.def} 
\end{eqnarray}

\begin{theorem}\label{T2}
Suppose condition of Theorem~\ref{T1} are satisfied. Let a symmetric
positive semidefinite $Nn\times Nn$ matrix $P=P'$ and a 
constant $\bar\gamma>0$ be such that there exit symmetric matrices $\bar
R_i> 0$, $\bar X_i> 0$, $\bar Z_{ij}>0$, $j\in 
\mathbf{V}_i$, $i=1, \ldots N$, such that  
\begin{enumerate}[(a)]
\item
The following matrix inequality is satisfied
\begin{equation}
\label{LMI.1}
  \bar R+\bar\gamma^2(\bar\Phi+\bar\Phi'-\bar\Delta) > P,
\end{equation}
where $\bar R=\mathrm{diag}[\bar R_1,\ldots,\bar R_N]$;  
\item
Each differential Riccati equation  
\begin{eqnarray}\label{Riccati.1} 
\hspace{-.5cm}\dot{\bar{Y}}_i &=& A(t)\bar Y_i+\bar Y_iA(t)'- \bar Y_i(C_i(t)'(D_i(t)D_i(t)')^{-1}C_i(t) 
\nonumber \\
\hspace{-.5cm}&+& \sum_{j\in\mathbf{V}_i} W_{ij}'\bar U_{ij}^{-1}W_{ij}
-\frac{1}{\bar\gamma^2}\bar R_i)\bar Y_i
+B_{1,i}(t)B_{1,i}(t)', \qquad  \\
\hspace{-.5cm}&&\bar Y_i(0)=\bar X_i^{-1}, \nonumber  
\end{eqnarray}
has a positive definite symmetric bounded solution $\bar Y_i(t)$ on the
interval $[0,\infty)$.  
\end{enumerate}
Then the network of systems (\ref{e}),  
 with the coefficients $L_i^r$, $K_{ij}^r$, obtained as
\begin{eqnarray}
  \label{L1}
  L_i^r(t)&=&\bar Y_i(t)C_i(t)'(D_i(t)D_i(t)')^{-1}, \nonumber \\
  K_{ij}^r(t)&=&\bar Y_i(t)W_{ij}'\bar U_{ij}^{-1},
\end{eqnarray}
has the following properties: 
\begin{enumerate}[(i)]
\item 
When $w=0$, $v_i=0$, and $v_{ij}=0$ for all $j\in\mathbf{V}_i$,
  $i\in \mathbf{V}$, and the system is not under attack, i.e., $f_i=0$ for
  all $i\in\mathbf{V}$, every error $e_i(t)$ of the
  distributed observer (\ref{e}) vanishes exponentially.
\item
In the presence of disturbances and/or when the system is subjected to a
biasing attack of class $\mathcal{F}_a$, $e_i(t)$ converge to 0
asymptotically. Furthermore, it holds that  
\begin{eqnarray}
\lefteqn{\int_0^{\infty}\!\!\mathbf{e}'P\mathbf{e}dt\le \bar\gamma^2
  \sum_{i=1}^N\bigg(\|x_0-\xi_i\|_{\bar X_i}^2} && 
\nonumber \\
&&+ \int_0^\infty\!\!\big(\|w\|^2+\|v_i\|^2+\|\varphi_i\|^2+\!\sum_{j\in
  \mathbf{V}_i}\|v_{ij}\|^2\big)dt\bigg). \qquad  
\label{z.Hinf.prime}
\end{eqnarray}
\end{enumerate}
\end{theorem}

\emph{Proof: }
First consider
the case where $w_i=0$, $v_i=0$, $v_{ij}=0$, and $f_i=0$. Note that the
latter assumption implies $\nu_i=0$ due to (\ref{eq:2}). Therefore, in this
case $w_i$ reduces to $w_i =[0'~ \tilde \nu_i']'$. Let $\Psi(t,\tau)$ be
the state transition matrix of the large scale system comprised of
subsystems (\ref{e.res}). Using Lemma~\ref{stabt}, we obtain that $\|\Psi(t,t_0)\|\le \beta_0
e^{-\lambda_0 (t-t_0)}$. Also, in Theorem~\ref{T1} we have established that
the signal $\tilde \nu(t)$ decays 
exponentially to 0 when $w_i=0$, $v_i=0$, $v_{ij}=0$, and $f_i=0$ for all $i$,
$j\in\mathbf{V}_i$.
Together these observations imply that the response of the system to
$\tilde \nu=[\tilde \nu_1'\ldots \tilde \nu_N']'$ vanishes exponentially,
i.e., claim (i) of the theorem holds. Statement (ii) follows directly from
Lemma~\ref{stabt}. \hfill$\Box$

\subsubsection*{Step 3. The complete controlled observer}

We now complete the last step of the design procedure and obtain the
remaining coefficients $\bar L_i^r$, $\bar K_{ij}^r$ for the controller
(\ref{ext.obs.nu.1.om.res}) from (\ref{LK}). 

\begin{theorem}\label{main}
Suppose the conditions of Theorems~\ref{T1} and~\ref{T2} are
satisfied. Let the coefficients $\hat L_i^r$, $\hat K_{ij}^r$, $\check
L_i^r$, $\check K_{ij}^r$ of the attack detecting controllers
(\ref{ext.obs.nu.1.om.res}) be obtained from (\ref{L}), and let 
$L_i^r$, $K_{ij}^r$ be the matrices defined in (\ref{L1}). Define 
$\bar L_i^r$, $\bar K_{ij}^r$ as
\begin{eqnarray}
  \label{barK}
  \bar L_i^r=\hat L_i^r-L_i^r, \quad \bar K_{ij}^r=\hat K_{ij}^r-K_{ij}^r. 
\end{eqnarray}
Then, the network of state observers (\ref{UP7.C.d.res}), augmented with
the network of attack detectors (\ref{ext.obs.nu.1.om.res})--(\ref{varphi})
produces state 
estimates $\hat x_i$ which have the following convergence properties:
\begin{enumerate}[(i)]
\item
In the absence of disturbances and biasing attacks, $\|x-\hat x_i\|\to 0$
exponentially as $t\to\infty$. 
\item
When the plant and/or the network is subject to $L_2$-integrable
disturbances and/or admissible biasing attacks, the estimates $\hat x_i$
converge to $x$ asymptotically as $t\to \infty$, and the resilient
performance of this observer network is characterized by the condition 
\begin{eqnarray}
\lefteqn{\int_0^{\infty}\mathbf{e}'P\mathbf{e}dt}
&& 
\nonumber \\
&&\le \bar\gamma^2
  \sum_{i=1}^N\bigg(\|x_0-\xi_i\|_{\bar X_i+2\gamma^2 X_i}^2 \nonumber \\
&&+ (1 + 2\gamma^2)\int_0^\infty\!\!\big(\|w\|^2+\|v_i\|^2+\sum_{j\in
    \mathbf{V}_{ij}}\|v_{ij}\|^2\big)dt\bigg) \nonumber \\
&&+ 2\bar\gamma^2 (1+\gamma^2) \sum_{i=1}^N
\int_0^\infty\|\nu_i\|^2dt;
\label{z.Hinf.final}
\end{eqnarray}
here $P$ is the matrix from condition (\ref{LMI.1}) of Theorem~\ref{T2}.
\item
Also, the outputs $u_i$ of the distributed controllers
(\ref{ext.obs.nu.1.om.res}) have the following properties:
\begin{enumerate}[(a)]
\item
If the node $i$ is not under attack, the signal $u_i$ generated by the
controller (\ref{ext.obs.nu.1.om.res}) at this node vanishes
asymptotically, even in the presence of disturbances.
\item
If node $i$ is subjected to a biasing attack $f_i$ of the class $\mathcal{F}_a$,
then $\int_0^\infty\|f_i-u_i\|^2dt<\infty$.
\end{enumerate}
\end{enumerate}
\end{theorem}

\emph{Proof: }
Statements (i) and (ii) of the theorem follow from Theorems~\ref{T1}
and~\ref{T2}. The inequality~(\ref{z.Hinf.final}) is proved by combining
(\ref{z.Hinf}) and (\ref{z.Hinf.prime}) using the inequality
$\|\varphi_i\|^2\le 2(\|\nu_i\|^2+\|\Upsilon_i \delta_i\|^2)$. 

To prove statement (iii)a, we note that if node $i$ is not under attack,
we have $\hat f_i(t)=0$ and $\nu_i(t)=0$, therefore
$\epsilon_i(t)=0$. Further, we have established in 
Theorem~\ref{T1} that in the presence of disturbances, $\delta_i(t)\to 0$
as $t\to\infty$ asymptotically. Therefore in this case,
$u_i=\Upsilon_i\hat\epsilon_i= -\Upsilon_i \delta_i$ vanishes to $f_i=0$ asymptotically as $t\to\infty$.  

When the observer at node $i$ is subjected to a biasing attack and $f_i\neq
0$, then $\nu_i\neq 0$, however $\nu_i\in L_2[0,\infty)$
according to Definition~\ref{def.admiss}. Also, it has been established in
Theorem~\ref{T1} that $\Upsilon_i \delta_i\in L_2[0,\infty)$. Thus,
$\int_0^\infty\|f_i-u_i\|^2dt<\infty$, i.e., (iii)b holds.  
\hfill$\Box$

\subsection{Performance optimization over communication graphs: an LTI case}\label{opt}

As mentioned, the proposed procedure allows each node to compute    
the coefficients $L_i^r$, $K_{ij}^r$ of its observers and the parameters
$\bar L_i^r$, $\bar K_{ij}^r$, $\check L_i^r$, $\check K_{ij}^r$ of
its controllers without communicating with other nodes. This is because each
node solves its Riccati equations (\ref{Riccati}) and (\ref{Riccati.1})
locally. For this, each node must be assigned with constants $\gamma^2$,
$\bar\gamma^2$ and matrices $\mathbf{R}_i$,
$\bar R_i$, $Z_{ij}$, $\bar Z_{ij}$. A deeper look into the
matrix inequalities (\ref{LMI}), (\ref{LMI.1}) reveals that the selection
of these matrices and constants is constrained by the network topology. The
simplest way to see this is to restrict attention to case where the
matrices $A$, $B$, $C_i$ and $D_i$ are constant, the
communications between the network nodes are 
noise-free, and the node observers transmit their complete estimates $\hat
x_i$. In this case, the 
communication model (\ref{communication}), (\ref{eq:eta}) simplifies to
$c_{ij}=\hat x_j$, $\eta_{ij}=\hat e_j$
for all $j\in\mathbf{V}_i$, $i=1,\ldots, N$; i.e., $W_{ij}=I$, $\hat H_{ij}=0$. Also, let $Z_{ij}=\bar Z_{ij}=I$. With these
parameters, we have  
\begin{equation}
  \label{eq:9}
  \Delta_i=\bar\Delta_i=d_iI, \quad \Phi=\bar\Phi=\mathcal{L}\otimes I,
\end{equation}
where $d_i$, $\mathcal{L}$ denote the in-degree of node $i$ and the Laplace
matrix of the network graph, respectively. Let $\mathcal{D}=\mathrm{diag}[d_1,\ldots,
d_N]$ be the in-degree matrix of the network graph. Conditions
(\ref{LMI}), (\ref{LMI.1}) can then be explicitly expressed in terms of the Laplace
and in-degree matrices of the graph $\mathbf{G}$: 
\begin{eqnarray}
  \label{eq:10}
  &&R+\gamma^2 (\mathcal{L}+\mathcal{L}'-\mathcal{D})\otimes I >0, \quad
  \check R_i> \Upsilon_i'\Upsilon_i, \nonumber \\
  &&\bar R+\bar\gamma^2 (\mathcal{L}+\mathcal{L}'-\mathcal{D})\otimes I > P.
\end{eqnarray}
These 
constraints can be used to select the 
communication topology which endows the controlled distributed
observer (\ref{UP7.C.d.res}), (\ref{ext.obs.nu.1.om.res})--(\ref{varphi})
with an optimized estimation accuracy or an optimized biasing attack 
detection performance.  

\subsubsection*{Optimization of the attack detection performance}
Let $\{\mathbf{G}_m, m=1, \ldots, M\}$ be a given
collection of admissible communication graphs. From now, we will use the
subscript $_m$ to denote quantities corresponding to the graph
$\mathbf{G}_m$ from this set; i.e., $\mathcal{L}_m$ will denote the Laplace
matrix of $\mathbf{G}_m$, etc. The optimized disturbance attenuation
among attack detectors (\ref{ext.obs.nu.1.om.res})--(\ref{varphi})
interconnected over the admissible graphs 
$\mathbf{G}_m$ is expressed as 
\begin{equation}
  \label{eq:12}
  \min_{\mathbf{G}_m} (\inf \gamma^2),
\end{equation}
where the infimum is taken over the set of matrices $\bar
R_i$, $R_i$, $\check R_i$, $i=1, \ldots, N$, and constants $\gamma^2$,
$\bar\gamma^2$ which satisfy the conditions 
of Theorem~\ref{main} stated for the graph $\mathbf{G}_m$.  
In the $H_\infty$ theory, a filter delivering 
optimal disturbance attenuation is usually difficult to compute,
and a standard practice is to use suboptimal $H_\infty$
filters~\cite{Basar-Bernhard,ZDG-1996}. Therefore, we
seek to find a resilient distributed observer with an optimized suboptimal
attack detection performance. 

Initially, let us restrict attention to the simplified
time-invariant case. In addition to the assumptions made above, we assume
that $(A,B)$ is stabilizable. With this additional assumption,
~\cite[Theorem~2]{ZU1a} states that the network of observers connected over
an admissible graph $\mathbf{G}_m$ satisfies the conditions of
Theorem~\ref{main} provided the following Linear Matrix Inequalities (LMIs) in the variables $(\{\bar R_i,R_i,\check R_i,\mathbf{Q}_i,\bar
Q_i\}_{i=1}^M,\gamma^2,\bar\gamma^2)$  are feasible:  
\begin{eqnarray}
&&R+\gamma^2 (\mathcal{L}_m+\mathcal{L}_m'-\mathcal{D}_m)\otimes I >0,\nonumber \quad
  \check R_i> \Upsilon_i'\Upsilon_i, \\ 
&& \bar R + \bar\gamma^2(\mathcal{L}_m+\mathcal{L}_m'-\mathcal{D}_m)\otimes
I> P, 
\nonumber \\
&&
\left[\begin{array}{c|c}\begin{array}{l}\mathbf{A}_i'\mathbf{Q}_i+\mathbf{Q}_i\mathbf{A}_i
      +\mathbf{R}_i\\
      - \left[
    \begin{array}{cc}
      \gamma^2(C_i'(D_iD_i')^{-1}C_i+d_{i,m}I) & 0 \\
 0 & 0
\end{array}\right]
\end{array}
& \mathbf{Q}_i\mathbf{B}_i\\[3ex] \hline
  \mathbf{B}_i'\mathbf{Q}_i & -\gamma^2I\end{array}\right]<0,  \nonumber \\
&&\left[\begin{array}{cc}\begin{array}{l}A'Q_i+Q_iA+\bar R_i\\-\bar\gamma^2\left(C_i'(D_iD_i')^{-1}C_i+d_{i,m}I\right)
\end{array}
& Q_iB_{1,i}\\
  B_{1,i}'Q_i & -\bar\gamma^2I\end{array}\right]<0,  \nonumber \\
  && \mathbf{Q}_i=\mathbf{Q}_i'>0, \quad R_i=R_i'>0,  
\nonumber \\
  && Q_i=Q_i'>0, \quad \bar R_i=\bar R_i'>0, \qquad
  (i=1,\ldots,N), \quad \label{eq:17}
\end{eqnarray}
where $d_{i,m}$ is the cardinality of $i$'s neighborhood $\mathbf{V}_i^m$
in the graph $\mathbf{G}_m$. 
Indeed, the first two
lines are the inequalities (\ref{eq:10}) particularized for the graph
$\mathbf{G}_m$, and the remaining conditions guarantee that the
corresponding differential Riccati equations (\ref{Riccati}),
(\ref{Riccati.1}) do not have a conjugate point~\cite{Basar-Bernhard}.  

The following sequential optimization
procedure provides a tractable upper bound on (\ref{eq:12}) which yields a
desired suboptimal attack detector.  Let $\gamma_0^\circ=+\infty$, and for
every $m=1,\ldots, M$, define $\gamma_m^2=\inf\gamma^2$ where the infimum is
taken over the  feasible set of the LMIs (\ref{eq:17}).
If these LMIs are not feasible for a particular $m$, we set
$\gamma_m^2=+\infty$. Then  let
\begin{eqnarray}
  \label{eq:13}
  \gamma_m^\circ = \min (\gamma_m, \gamma_{m-1}^\circ),
  \quad m=1, \ldots, M.
\end{eqnarray}
Since the sequence
$\{\gamma^\circ_m\}$ is monotone decreasing,  the recursion
(\ref{eq:13}) terminates at the graph $\mathbf{G}_m$ which attains  
$\min_{\mathbf{G}_m}(\gamma_m^\circ)^2=\min_{\mathbf{G}_m}\gamma_m^2$. Also,
since feasibility of the LMIs (\ref{eq:17}) is only a sufficient condition
for the conditions of Theorem 3 to be satisfied, then for every $m$,
$\gamma_m^2$ is greater than or equal to the inner infimum 
value in (\ref{eq:12}). Therefore, the graph $\mathbf{G}_m$ which attains 
$\min_{\mathbf{G}_m}(\gamma_m^\circ)^2$ is the most favorable
graph among the candidate graphs $\{\mathbf{G}_m\}$, from the view point of   
suboptimal attack detection performance.    

\subsubsection*{Optimization of resilient estimation performance}
The foregoing procedure is readily modified to obtain a
network topology yielding a suboptimal level of disturbance
attenuation $\bar\gamma^2$ in (\ref{z.Hinf.prime}). Define
$\bar\gamma_m^2= \inf\bar\gamma^2$ where the infimum is
taken over the  feasible set of the LMIs (\ref{eq:17}). Then the recursion 
\begin{eqnarray}
  \label{eq:13bar}
  \bar\gamma_m^\circ = \min (\bar\gamma_m, \bar\gamma_{m-1}^\circ),
  \quad m=1, \ldots, M.
\end{eqnarray}
terminates at the graph which attains the desired suboptimal network
configuration.    

\subsubsection*{General time-invariant case}
The optimization procedures proposed above
are applicable in the general case where the communications between the
nodes are described
by the general model (\ref{communication}), (\ref{eq:eta}). We still assume
that the 
matrices $Z_{ij}$, $\bar Z_{ij}$ are given and that the pair $(A,B)$ is
stabilizable. Then $\gamma^2$ and $\bar \gamma^2$ can be optimized 
subject to a series of linear matrix inequalities, similar to
(\ref{eq:17}). In lieu of the first three inequalities in (\ref{eq:17}),
this series of LMIs includes the original inequalities 
(\ref{LMI}) and (\ref{LMI.1}) involving
the appropriately defined matrices $\Phi_m+\Phi_m-\Delta_m$ and
$\bar\Phi_m+\bar\Phi_m-\bar\Delta_m$ which are associated
with the admissible candidate network topology $\mathbf{G}_m$. Also, the
remaining matrix inequalities in (\ref{eq:17}) 
are replaced with the corresponding more general LMIs reflecting the
general structure of communications
\begin{eqnarray}
&& \hspace{-2ex} \left[\begin{array}{cc}{\small\begin{array}{l}\mathbf{A}_i'\mathbf{Q}_i+\mathbf{Q}_i\mathbf{A}_i+\mathbf{R}_i\\-\gamma^2\left(\mathbf{C}_i'(D_iD_i')^{-1}\mathbf{C}_i+\sum\limits_{j\in\mathbf{V}_i^m}\mathbf{W}_{ij}'U_{ij}^{-1}\mathbf{W}_{ij}\right)
\end{array}}
& \mathbf{Q}_i\mathbf{B}_i\\
  \mathbf{B}_i'\mathbf{Q}_i & -\gamma^2I\end{array}\right]<0, \nonumber \\
&& \hspace{-2ex} \left[\begin{array}{cc}{\small\begin{array}{l}A'Q_i+Q_iA+\bar R_i\\-\bar\gamma^2\left(C_i'(D_iD_i')^{-1}C_i+\sum\limits_{j\in\mathbf{V}_i^m}W_{ij}'U_{ij}^{-1}W_{ij}\right)
\end{array}}
& Q_iB_{1,i}\\
  B_{1,i}'Q_i & -\bar\gamma^2I\end{array}\right]<0, \nonumber \\
&& \mathbf{Q}_i=\mathbf{Q}_i'>0, \quad R_i=R_i'>0,  
\nonumber \\
&& Q_i=Q_i'>0, \quad \bar R_i=\bar R_i'>0, \qquad
  (i=1,\ldots,N), \quad \label{eq:17a}
\end{eqnarray}
The optimization over the set of graphs can then be
performed recursively, in the same way as in the previous cases.  

\subsubsection*{Suitable candidate graphs}
Our final remarks are concerned with selecting suitable candidate graphs
$\mathbf{G}_m$ for optimization. 
   
The feasibility of the proposed optimization procedure relies on
$H_\infty$ stabilizability of the large-scale interconnected systems
(\ref{ext.error}) and (\ref{e.res}) via output injection. In turn, this
stabilizability property requires the
corresponding noise-free large-scale systems to have a basic
distributed detectability property; see~\cite{U7b-journal} for a more
detailed discussion. For observer networks 
with identical matrices $W_{ij}=W$, this property requires that the
plant must be detectable from the combined outputs associated with
each maximal subgraphs spanned by a tree~\cite{U7b-journal} or with each 
strongly connected subgraph that does not have incoming
edges~\cite{PM-2017,MS-2018,WM-2017}. In addition, within each such
subgraph, every state of the plant must be either detectable from the
measurements or observable through interconnections (or
both)\footnote{Observability through interconnections 
  holds trivially in~\cite{PM-2017,WM-2017,HTWS-2018} since these
  references assume $W=I$. In~\cite{MS-2018}, the matrices $W_{ij}$ were
  selected to ensure that each node receives from its neighbors the part
  of the state vector undetectable from its
  local measurements.}. 
Although we did not state these properties explicitly in this paper, they
are necessary for stabilizability of the large-scale 
interconnected systems (\ref{ext.error}) and (\ref{e.res}) via output
injection. In particular, a necessary condition for detectability of
biasing attacks by detectors of the form (\ref{ext.obs.nu.1.om.res})
obtained in~\cite{DUSL1} is based on these properties. 
Therefore, each candidate graph $\mathbf{G}_m$ must satisfy
these necessary conditions for stablizability via output injection, at
least when the system is time-invariant, and 
the matrices $W_{ij}$ are identical\footnote{The extension of the results
  of~\cite{U7b-journal} for the case where $W_{ij}=W_i$ can be found
  in~\cite{U7c}.}. We are not aware of similar necessary conditions
for distributed detectability of time-varying systems, and in this case the
problem of characterizing distributed detectability as well as detectability  
of biasing attacks appears to remain open.

\section{Illustrative example}\label{sec:simulations}
The example is based on the example from \cite{U6}, where a distributed
observer was constructed for the system of the form (\ref{eq:plant}), with
$B = 0.1I$ and 
\begin{eqnarray*}
&&A = \left[\begin{array}{cccccc}
  0.3775 &      0 &      0 &        0 &      0 &       0 \\
  0.2959 & 0.3510 &      0 &        0 &      0 &       0 \\
  1.4751 & 0.6232 & 1.0078 &        0 &      0 &       0 \\
  0.2340 &     0  &      0 &   0.5596 &      0 &       0 \\
       0 &      0 &      0 &   0.4437 & 1.1878 & -0.0215 \\
       0 &      0 &      0 &        0 & 2.2023 &  1.0039 \\
\end{array}\right].
\end{eqnarray*}
The plant is observed by 6 sensors. Sensor $i$ measures the $i$-th and
$(i+1)$-th coordinates of the 
state vector, with sensor 6 measuring the 6th and 1st
coordinates; see~\cite{U6} for the definitions of the matrices $C_i$. All 6
pairs $(A,C_i)$ are not detectable in this example. Also,
$D_i = 0.01 I$ $\forall i$. As in~\cite{U6}, suppose that the nodes
broadcast the full vector $\hat x_i$, i.e., $W=I$, however the
communications between the nodes are subject to disturbances, so we let 
$H_{ij}=H_{c,ij}=(0.1/\sqrt{2})\times[1~1~1~1~1~1]'$.

The problem in~\cite{U6} was to achieve a robust $H_\infty$ consensus
performance of the observers, which corresponds to 
$P=(\mathcal{L}+\mathcal{L}_{\mathrm{T}})\otimes I$; here $\mathcal{L}$, $\mathcal{L}_{\mathrm{T}}$ are the
Laplace matrices of $\mathbf{G}$ and its transpose
graph, respectively. We now consider
a resilient version of that problem. As in~\cite{DUSL1}, suppose that 
$F_i=[1~1~1~1~1~1]'$, $\hat F_i=0$ $\forall i$. 
Next we selected $G_i(s)=\frac{410}{s+40}$. 
Formally, the algorithm
imposes mild requirements on these transfer 
functions --- any proper rational transfer function of the form
$\frac{N(s)}{D(s)}I$ which renders the transfer function $(sI+G_i(s))^{-1}G(s)$
    stable can be selected as $G_i(s)$. Some additional considerations in
    regard to selecting these transfer functions 
    are as follows. Firstly, it is reasonable to keep the energy in
    the approximation error $\nu_i$ to a minimum if possible, since it appears
    on the right hand side of the performance
    inequality~(\ref{z.Hinf.prime}). For this, the 
    system (\ref{eq:2}) should be sufficiently fast and have no
    overshoots. Secondly, the differential Riccati equation
(\ref{Riccati}) must admit a bounded solution with an acceptably small
$\gamma$, to guarantee an acceptable robustness of attack detection; see
(\ref{delta.Hinf.i}). This can be accomplished by either simulating those
equations offline or, in the time-invariant case, by testing that the third
LMI condition~(\ref{eq:17}) is feasible. In accordance with these
recommendations, in this example we let $Z_{ij}=\bar
Z_{ij}=0.01\times I$ and solved the LMI 
conditions~(\ref{eq:17}) to obtain the smallest $\gamma^2$,
$\bar\gamma^2$ for which the LMIs~(\ref{eq:17}) were feasible,
$\gamma^2= 6.9113\times 10^{-3}$, $\bar\gamma^2= 3.4511\times 10^{-2}$, and
obtained the corresponding matrices $R_i$, $\check R_i$ and $\bar R_i$
from~(\ref{eq:17}) to be used in the Riccati equations (\ref{Riccati}) and
(\ref{Riccati.1}). To be consistent with the original example from~\cite{U6}, we
consider only one network topology, hence we did not need to perform
optimization over a collection of graphs.
  
\begin{figure} [t]
\centering
    \begin{subfigure}[t]{\columnwidth}
  \includegraphics[width=0.8\columnwidth,height=4.8cm]{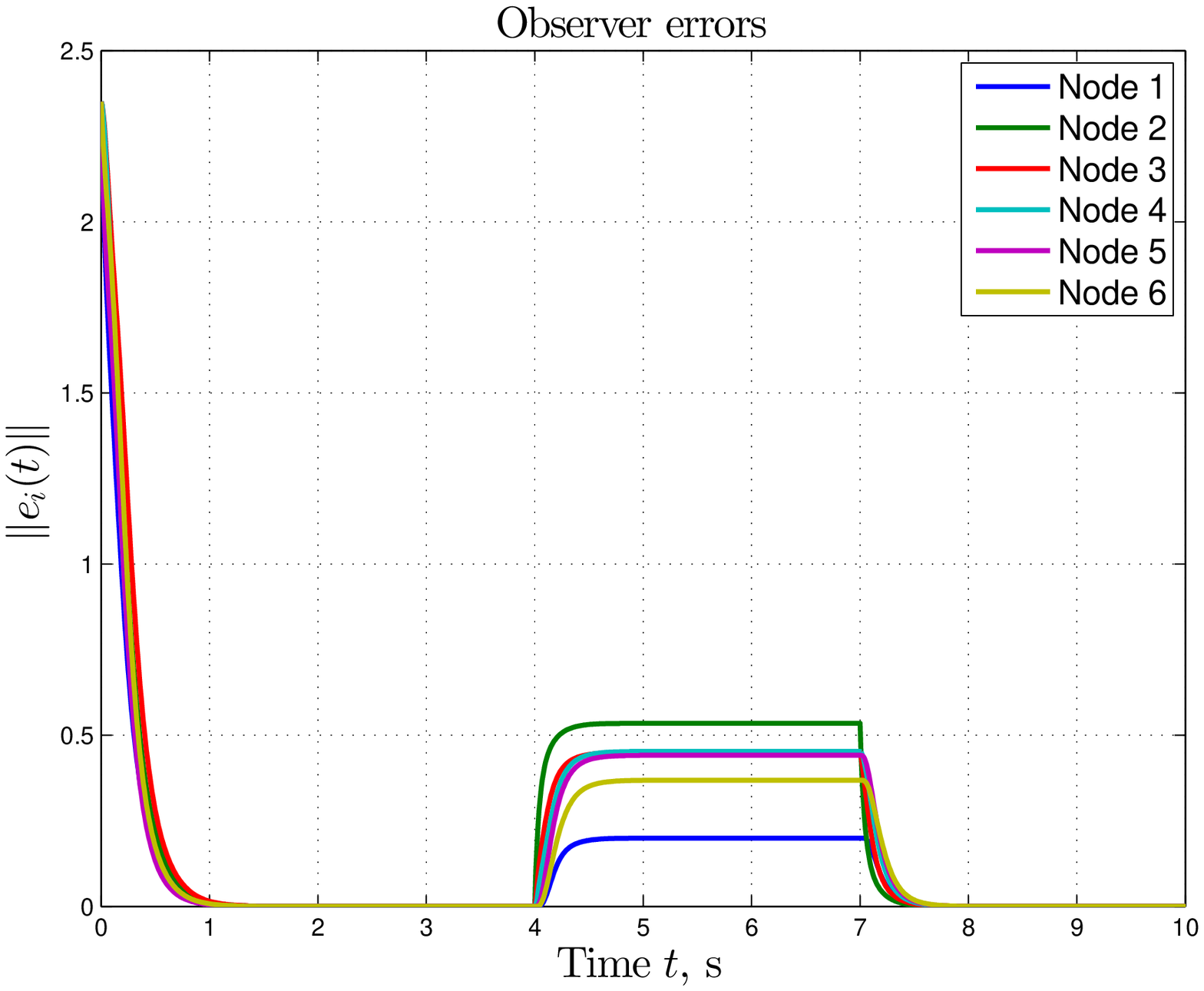} (a)
    \end{subfigure}
    \begin{subfigure}[t]{\columnwidth}
  \includegraphics[width=0.8\columnwidth,height=4.8cm]{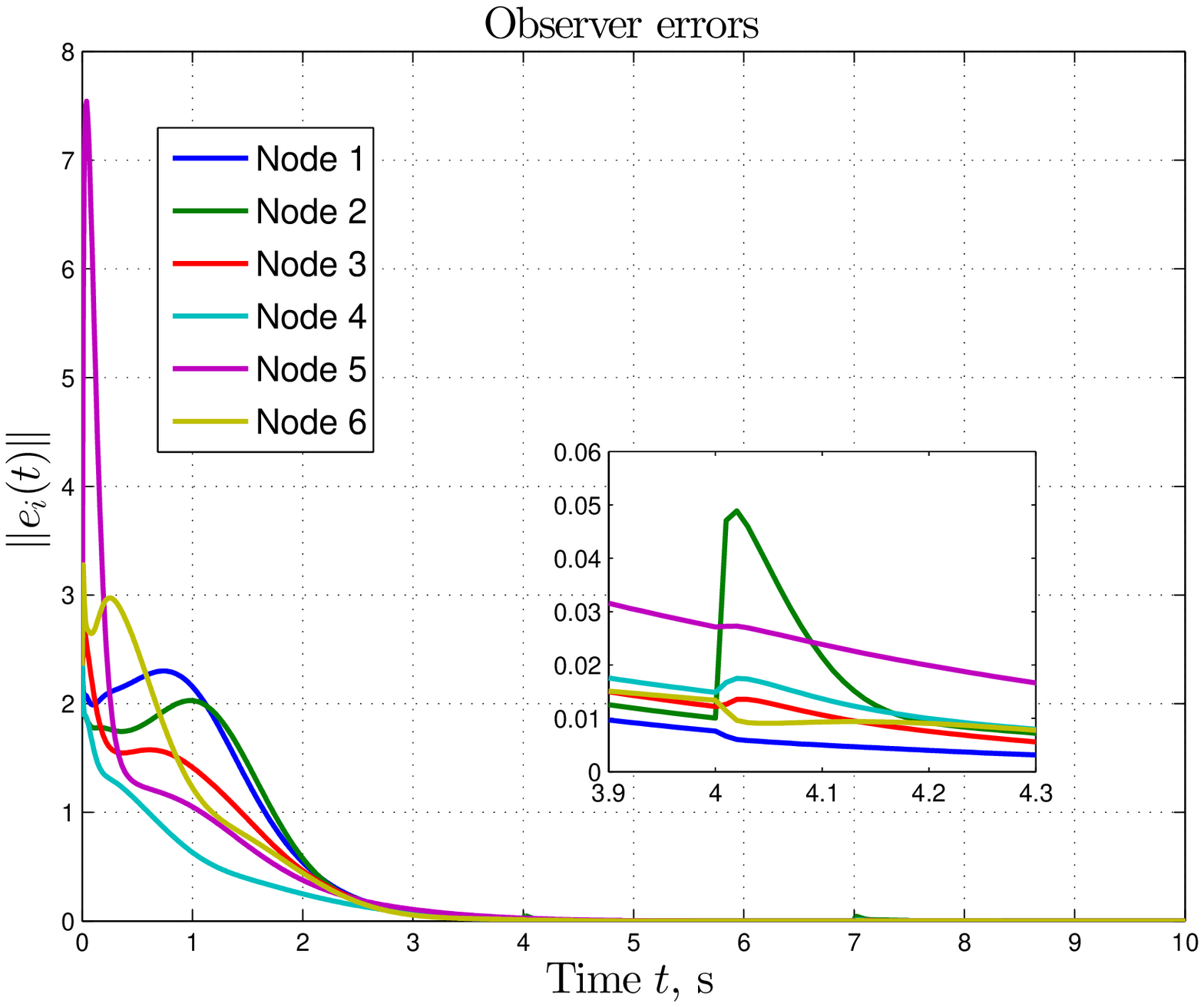} (b)
    \end{subfigure}
  \caption{Errors of the original observers (\ref{UP7.C.d.unbiased}) 
    from~\cite{U6} (Fig.(a)) and the controlled observers (\ref{UP7.C.d.res})
    (Fig.(b)) under the biasing attack. The inset shows the zoom-up plots of
    the transients in the interval $t\in[3.9,4.3]$.}  
  \label{fig:f2}  \label{fig:f3}
\end{figure}

\begin{figure} [t]
\centering
  \includegraphics[width=0.8\columnwidth,height=4.8cm]{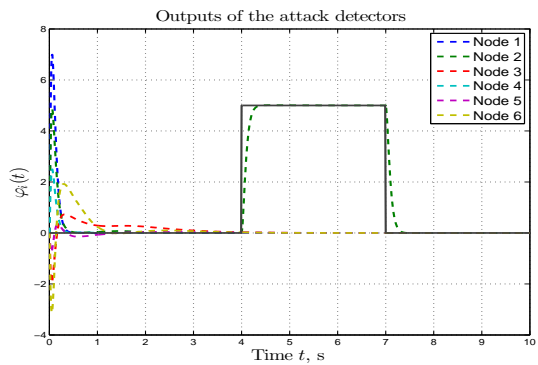}
  \caption{Outputs of the attack detectors $\varphi_i(t)$. The solid line
    shows the attack input, and the colored dashed lines show the detector 
    outputs. } 
  \label{fig:f1}
\end{figure}

Next, the plant and the resilient observers (\ref{UP7.C.d.res}) endowed
with the controllers (\ref{ext.obs.nu.1.om.res})--(\ref{varphi}) were jointly
simulated. An attack input of amplitude 5 and duration 3 seconds was applied to
the plant observer at node 2 at time 
$t=4$. No other disturbances were applied; this
made the comparison between the errors of the original
observer (\ref{UP7.C.d.unbiased}) from~\cite{U6} and the errors of the observers
(\ref{UP7.C.d.res}) operating under the attack most vivid; see
plots in Fig.~\ref{fig:f2}. As can be seen from these plots, the
observers from~\cite{U6} designed without consideration for resilience were
adversely affected by the attack. In contrast, the controlled
observers (\ref{UP7.C.d.res}) were able to
successfully negate the biasing effect of the attack. Their errors have only short transients at the
beginning and the end of the attack interval, which decay quite quickly; the decay
rate was adjusted by choosing the transfer functions
$G_i(s)$. Figure~\ref{fig:f1} shows the corresponding outputs $u_i$ of the
controllers. One can see that the attack at node 2 has been detected
successfully.      

      
\section{Conclusion} \label{sec:conclusion}
The paper has proposed a novel class of controlled distributed
observers for robust estimation of a linear plant, which are resilient 
to biasing misappropriation attacks. To counter the attacks, we
introduce an auxiliary network of distributed output feedback controllers
which provide an attack correcting action to the main observer. The filter
components of these controllers can also serve as attack detectors. Our
problem formulation is quite general in that it captures biasing attack
scenarios which target both the observer and the attack detector at the
compromised nodes.      

The proposed design method is based on the methodology of distributed
$H_\infty$ filtering; it allows to obtain 
observers which attenuate benign disturbances, while discovering and
suppressing biasing inputs. This approach allows to consider the
worst-case situation where every node of the observer network can be subjected
to an attack. When the defender knows \emph{a priori} that
certain nodes are secure (e.g., as a result of a security audit), this
information can be easily incorporated into the design procedure by
assigning zero values to the corresponding attack input matrices $F_i$,
$\check F_i$. 

Another feature of our approach is that the nodes compute their
observers and attack detectors independently from each other. This
decentralization of computation enhances security of the network since the
computation does not rely on potentially
vulnerable communications. 

\section{Acknowledgment}
The author is grateful to the anonymous referees whose comments helped to
improve the presentation of the paper.

\newcommand{\noopsort}[1]{} \newcommand{\printfirst}[2]{#1}
  \newcommand{\singleletter}[1]{#1} \newcommand{\switchargs}[2]{#2#1}

\end{document}